\documentclass[aps,prc,preprint, nofootinbib, groupedaddress]{revtex4-1}
\usepackage{amsmath}
\usepackage{mathtools}
\usepackage{braket}
\usepackage{multirow}
\usepackage{amssymb}
\usepackage[x11names]{xcolor}
\usepackage[normalem]{ulem}

\begin{document}
\title{Incoherent solar-neutrino scattering off the stable Tl isotopes}

\author{Dimitrios K. Papoulias}
\affiliation{Department of Physics, National and Kapodistrian University
of Athens, Zografou Campus GR-15772 Athens, Greece}
\email[]{dkpapoulias@phys.uoa.gr}
\author{Matti Hellgren}
\affiliation{University of Jyv\"askyl\"a, Department of Physics, P.O. Box 35, FI-40014, Finland}
\email[]{majokahe@jyu.fi}
\author{Jouni Suhonen}
\affiliation{University of Jyv\"askyl\"a, Department of Physics, P.O. Box 35, FI-40014, Finland}
\affiliation{International Centre for Advanced Training and Research in Physics (CIFRA), P.O. Box MG12, 077125, Bucharest-Magurele, Romania}
\email[]{jouni.t.suhonen@jyu.fi}

\begin{abstract}
Nuclear-structure calculations for the description of low-energy neutral-current neutrino scattering off the stable $^{203,205}$Tl isotopes are performed in the context of the nuclear shell model using the model space jj56pn. Cross-section and event-rate calculations focusing on inelastic solar-neutrino scattering off $^{203,205}$Tl are performed. The individual contributions of the various nuclear responses are presented and discussed, and the results are also illustrated in terms of the nuclear recoil energy. Analytical expressions entering the cross sections are given in order to achieve a direct connection with experimental observables.
\end{abstract}

\maketitle

\section{Introduction}

Neutrinos are among the most elusive particles in the standard model (SM), with little being known concerning their nature and fundamental properties. Their interactions with other particles proceed only via the weak interaction, thus the interactions have tiny cross sections. Neutrinos also interact with nuclei, with the corresponding cross sections being the largest in the neutrino sector. By  exploiting pion-decay-at-rest neutrinos, the first observation of coherent elastic neutrino-nucleus scattering (CE$\nu$NS) was reported by the COHERENT Collaboration in 2017 using a 14.6~kg CsI[Tl] detector~\cite{COHERENT:2017ipa}, and it was later confirmed in 2020 using a 24~kg liquid argon detector~\cite{COHERENT:2020iec}. These two measurements are consistent with the SM at $6.7\sigma$ and  $3.5\sigma$ confidence levels, respectively. In 2021 a new measurement on CsI[Tl] was reported with an improved background determination and double statistics which further rejected the no-CE$\nu$NS hypothesis at $11.6\sigma$~\cite{COHERENT:2021xmm}. Very recently, the COHERENT Collaboration reported a new low-statistics measurement on Ge which, however, reached a notable $3.9\sigma$ consistency with a SM CE$\nu$NS excess~\cite{Adamski:2024yqt}. Moreover,  suggestive evidence for CE$\nu$NS observation using a reactor has been reported by the Dresden-II Collaboration using a Ge detector~\cite{Colaresi:2022obx}. Finally, the first CE$\nu$NS observation  induced by $^8$B solar neutrinos was announced in July 2024, independently by PandaX-4T~\cite{PandaX:2024muv} and XENON~\cite{XENON_CEvNS} Collaborations  with a 2.7$\sigma$ significance, motivating further our present work. The field is very active, with several ongoing experiments aiming to detect neutrino-nucleus scattering events such as CONNIE~\cite{CONNIE:2021ggh}, CONUS~\cite{CONUS:2021dwh}, $\nu$GEN~\cite{nuGeN:2022bmg},
MINER~\cite{MINER:2016igy}, RICOCHET~\cite{Billard:2016giu}, NUCLEUS~\cite{Strauss:2017cuu}, TEXONO~\cite{Wong:2015kgl}, vIOLETA~\cite{Fernandez-Moroni:2020yyl}, RED-100~\cite{Akimov:2022xvr}, NEON~\cite{NEON:2022hbk}, NEWS-G~\cite{NEWS-G:2021mhf} and
the Scintillating Bubble Chamber (SBC)~\cite{SBC:2021yal} (for a recent review see~\cite{Abdullah:2022zue}).

From the theoretical point of view, the main source of uncertainty in the neutrino-nucleus scattering cross section comes from nuclear-physics effects. In this direction there have been a number of works exploring the nuclear-physics aspects of the CE$\nu$NS  process, following a phenomenological approach~\cite{Papoulias:2019lfi, Sierra:2023pnf}, shell model and Skyme-Hartree-Fock~\cite{AbdelKhaleq:2024hir}, deformed shell model~\cite{Kosmas:2021zve}, BCS~\cite{Papoulias:2015vxa}, Hartree-Fock + BCS~\cite{Co:2020gwl},  coupled-cluster~\cite{Payne:2019wvy} and shell model + chiral effective field theory~\cite{Hoferichter:2020osn}. While there have been intense efforts in probing new physics using CE$\nu$NS (see, e.g.,~\cite{Abdullah:2022zue} and references therein)  there are a limited number of studies focusing on the subdominant inelastic channels of neutral-current neutrino-nucleus scattering~\cite{Bednyakov:2018mjd,Sahu:2020kwh,Dutta:2022tav}. The latter have also been studied in the context of advanced nuclear-structure methods such as the quasiparticle random-phase approximation~\cite{Chasioti:2009fby, Tsakstara:2011zzc, Divari:2012cj,Almosly2015} and the microscopic quasiparticle-phonon model~\cite{Ydrefors:2011zza,Ydrefors2012,Hellgren:2022yvo}.

In this work we perform the required nuclear-structure calculations within the framework of the nuclear shell model. We mainly consider solar neutrinos for which we expect the bulk of the inelastic scattering contributions to arise from nuclear final states in the range 0--4~MeV of excitation. Nevertheless, we do not neglect states higher in energy and  also include final states up to 10--20~MeV. Despite the focus being on solar neutrinos, we also report the inelastic cross sections as functions of the energy of the incoming neutrino within the range 0-20~MeV. This information can be utilized by researchers in obtaining theoretical estimates of folded cross sections for low-energy neutrino sources beyond the solar neutrinos.

For the nuclei considered in this work, we specifically focus on the stable thallium isotopes $^{203,205}$Tl and present the expected solar-neutrino-induced event rates for both  CE$\nu$NS and inelastic neutrino-nucleus scattering channels. Thallium isotopes are of key interest since they constitute the dopant material of several detectors based on CsI[Tl] and NaI[Tl] crystals. Apart from COHERENT, thallium-doped materials are of particular interest in dark-matter direct detection searches and are currently in use by several experimental collaborations such as COSINE~\cite{COSINE-100:2019lgn}, DAMA/LIBRA~\cite{DAMA:2008jlt}, PICO-LON~\cite{PICO-LON:2015rtu}, ANAIS~\cite{Amare:2019jul}, and SABRE~\cite{SABRE:2018lfp}. An important novelty in the present work is the inclusion of nuclear recoil energy in the calculations.  To this purpose we  present, for the first time, the standard 
neutrino-scattering formalism in terms of the nuclear recoil energy for a more convenient comparison of the results with experimental data.

The paper is organized as follows: in Sec.~\ref{sec:formalism} we discuss the details of the nuclear-structure calculations performed and we also present the standard formalism to describe inelastic neutrino-nucleus scattering as well as  introduce the new formalism in terms of the nuclear recoil energy. Next, in Sec.~\ref{sec:results}, we discuss our main results and finally, in Sec.~\ref{sec:conclusions}, we highlight our concluding remarks.

\section{Theoretical Formalism}
\label{sec:formalism}
\subsection{Nuclear-structure calculations}
\label{sec:shell-model}
The nuclear many-body framework utilized in this paper for the nuclear-structure calculations is the nuclear shell model. For the stable thallium isotopes it is possible to use this model to derive the needed nuclear wave functions in large-scale computations. These wave functions are, in turn, used to construct the transition matrix elements between the ground state (g.s.) and the excited final states of the target nucleus. The calculations were conducted by utilizing the shell-model code NuShellX$@$MSU \cite{Brown:2014bhl} using the interaction \textit{khhe} within the model space jj56pn without any restrictions or truncations. All states of both parities from $J = 3/2$ to $J = 11/2$, along with the states with $J^{\pi} = 1/2^+$, were initially included. The $1/2^-$ states were excluded since the valence space does not contain like-nucleon orbitals that would differ in total angular momentum by 0 or 1 units and have opposite parity, so these states cannot be reached through a transition from the g.s. ($J^{\pi} = 1/2^+$ for both nuclei). Such transitions would be possible by including orbitals above or below the shell gaps, but this was judged to be unnecessary for the purposes of this work due to the hight excitation energy of these states and the low energy of solar neutrinos, leading to an expected negligible contribution of the $1^-$ multipole to the scattering cross section.

\begin{figure*}%[h][width = 0.49\textwidth, keepaspectratio]
\includegraphics{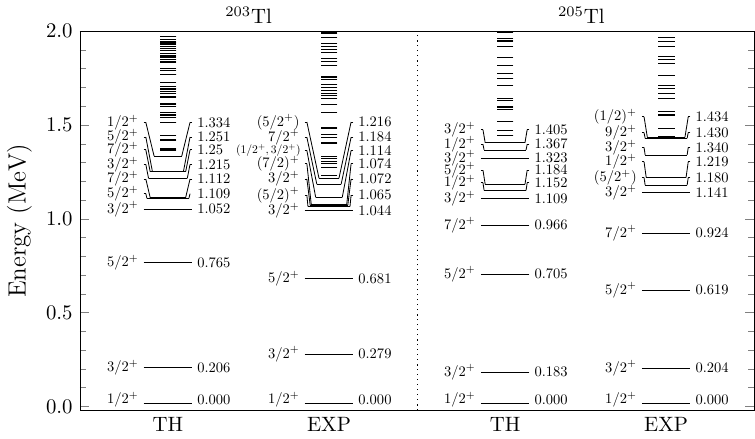}
\caption{Shell-model-computed spectra compared to the corresponding experimental spectra \cite{Kondev:2005bkq,Kondev:2020yee} for the nuclei of interest.}
\label{fig:levels}
\end{figure*}

For $^{205}$Tl there were a total of 1393 final states in the chosen model space, whereas for $^{203}$Tl that number was nearly 100000. To alleviate the computational burden of calculating the cross sections for scatterings off $^{203}$Tl, the number of final states included in the calculations was reduced considerably, i.e., we selected the first 1000 most contributing states. We  also verified that, under this approximation, our cross section results agree up to two decimal points with the full calculation assuming the complete set of final states. The details of this procedure are discussed in the following section. Some of the lowest states of the obtained spectra of the two nuclei, along with the corresponding experimental spectra for comparison, are illustrated in Fig.~\ref{fig:levels}.
The agreement between theory and experiment in the low-energy end of the spectra is good for both nuclei of interest, which is particularly important in low-energy solar-neutrino scattering where most of the contributions come from a small number of the low-energy states.

\subsection{Scattering cross section}
\label{sec:formalism_costeta}
\begin{figure*}%[h][width = 0.49\textwidth, keepaspectratio]
\includegraphics{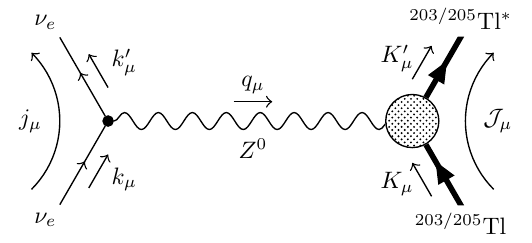}
\caption{Reaction kinematics illustrated in the lowest order Feynman diagram for the processes of interest. An incoming solar neutrino $\nu_e$ (4-momentum $k_{\mu}$) scatters off the nucleus $^{203}$Tl or $^{205}$Tl (4-momentum $K_{\mu}$), which is in its ground state prior to the scattering. The nucleus is left in an excited state denoted by an asterisk. The processes are mediated by an exchange of a $Z^0$ boson and the 4-momentum transfer is $q_{\mu} = k_{\mu} - k'_{\mu} = K'_{\mu} - K_{\mu}$.}
\label{fig:Feynman-diagram}
\end{figure*}

The scattering cross sections as functions of the energy of the incoming neutrino were calculated by using the standard Donnelly-Walecka formalism. The semileptonic nuclear processes
\begin{equation}
	\nu_{e} + \prescript{203/205}{}{\textrm{Tl}}(\textrm{ground state}) \quad \longrightarrow \quad \nu_{e} + \prescript{203/205}{}{\textrm{Tl}}^{\ast}(\textrm{excited state})
\end{equation}
proceed via an exchange of the neutral $Z^0$ boson, illustrated in the lowest order by a Feynman diagram of Fig.~\ref{fig:Feynman-diagram}.
This complicated process can be simplified by neglecting the propagator of the intermediate vector boson and treating the scattering as a point-like current-current interaction with the effective Hamiltonian
\begin{equation}
	\hat{H}_{\textrm{eff}} = \frac{G}{\sqrt{2}}\int\textrm{d}^3\textbf{x} \, j_{\mu} (\textbf{x})  \mathcal{J}^{\mu}(\textbf{x}),
\end{equation}
where $G \equiv G_F= 1.1664 \times 10^{-5}$ GeV is the Fermi constant and ($j_{\mu}$) $\mathcal{J}_{\mu}$ is the (lepton) hadron current. This approximation is appropriate when the transferred 4-momentum $q_{\mu} = k_{\mu} - k'_{\mu} = K'_{\mu} - K_{\mu}$ is sufficiently small. Here ($k_{\mu}$/$k_{\mu}'$) $K_{\mu}$/$K_{\mu}'$ is the 4-momentum of the initial/final state (neutrino) nucleus. The matrix element of the effective Hamiltonian connecting the initial ($i$) and final $(f)$ states can be written as
\begin{equation}
	\bra{f}\hat{H}_{\textrm{eff}}\ket{i} = \frac{G}{\sqrt{2}}\int\textrm{d}^3\textbf{x}\, e^{-i\textbf{q}\cdot\textbf{x}}l_{\mu}\bra{f}\mathcal{J}^{\mu}(\textbf{x})\ket{i},
\end{equation}
where we denote
\begin{equation}
	\bra{f}j_{\mu}(\textbf{x})\ket{i} \equiv e^{-i\textbf{q}\cdot\textbf{x}}l_{\mu}, \quad (l_0,\textbf{l}) = l_{\mu}
\end{equation}
for the lepton matrix element.

The differential cross section is related to the Hamiltonian matrix element through Fermi's golden rule \cite{Sakurai:2011zz}
\begin{equation}
	\frac{\textrm{d}\sigma}{\textrm{d}\Omega} = \frac{|\textbf{k}'|E_{\nu'}}{(2\pi)^2}V^2|\bra{f}\hat{H}_{\textrm{eff}}\ket{i}|^2,
\end{equation}
where ($\textbf{k}$) $\textbf{k}'$ and ($E_{\nu}$) $E_{\nu'}$ are the 3-momentum and total energy of the (initial-) final-state neutrino respectively, and $V$ is the quantization volume when the lepton fields are treated as box-normalized plane waves. The latter is related to the Dirac spinors by
\begin{equation}
	Vl_{\mu} = \overline{u}(\textbf{k}')\gamma_{\mu}(1-\gamma_5)\gamma_{\mu}u(\textbf{k}).
\end{equation}
In the case of unobserved and unpolarized targets the matrix-element part of the differential cross section can shown to be \cite{Donnelly:1978tz, Walecka:1995mi}
%\begin{equation} \label{eq:xsec_general} \begin{split}  &\frac{1}{2J_i + 1}\sum_{M_i,M_f}|\bra{f}\hat{H}_{\textrm{eff}}\ket{i}|^2 = \\& \frac{G^2}{2}\frac{4\pi}{2J_i + 1}\left\lbrace\sum_{J\geq1}\left[\frac{\textbf{l}\cdot\textbf{l}^* - l_3l_3^*}{2}\left(|\bra{J_f}\hat{\mathcal{T}}^{\textrm{mag}}_J(q)\ket{J_i}|^2 + |\bra{J_f}\hat{\mathcal{T}}^{\textrm{el}}_J(q)\ket{J_i}|^2\right) \right.\right. \\& \left.\left. - i\frac{\textbf{l}\times\textbf{l}^*}{2}\left(2\textrm{Re}\bra{J_f}\hat{\mathcal{T}}^{\textrm{mag}}_J(q)\ket{J_i}\bra{J_f}\hat{\mathcal{T}}^{\textrm{el}}_J(q)\ket{J_i}^*\right)\right] + \sum_{J\geq0}\left[l_3l_3^*\left(|\bra{J_f}\hat{\mathcal{L}}_J(q)\ket{J_i}|^2 \right.\right.\right. \\& \left.\left.\left. + l_0l_0^*|\bra{J_f}\hat{\mathcal{M}}_J(q)\ket{J_i}|^2\right) - 2\textrm{Re}\left(l_3l^*_0\bra{J_f}\hat{\mathcal{L}}_J(q)\ket{J_i}\bra{J_f}\hat{\mathcal{M}}_J(q)\ket{J_i}^*\right)\right]\right\rbrace,	\end{split} \end{equation}

\begin{equation} 
\label{eq:xsec_general} 
\begin{split}  
&\frac{1}{2J_i + 1}\sum_{M_i,M_f}|\bra{f}\hat{H}_{\textrm{eff}}\ket{i}|^2 = \\& \frac{G^2}{2}\frac{4\pi}{2J_i + 1} \Biggl\{\sum_{J\geq1}\left[\frac{\textbf{l}\cdot\textbf{l}^* - l_3l_3^*}{2}\left(|\bra{J_f}\hat{\mathcal{T}}^{\textrm{mag}}_J(q)\ket{J_i}|^2 + |\bra{J_f}\hat{\mathcal{T}}^{\textrm{el}}_J(q)\ket{J_i}|^2\right) \right. \\& \left.\left. - i\frac{\textbf{l}\times\textbf{l}^*}{2} 2\textrm{Re}\left(\bra{J_f}\hat{\mathcal{T}}^{\textrm{mag}}_J(q)\ket{J_i}\bra{J_f}\hat{\mathcal{T}}^{\textrm{el}}_J(q)\ket{J_i}^*\right)\right] + \sum_{J\geq0}\left[l_3l_3^*|\bra{J_f}\hat{\mathcal{L}}_J(q)\ket{J_i}|^2 \right.\right. \\& \left. + l_0l_0^*|\bra{J_f}\hat{\mathcal{M}}_J(q)\ket{J_i}|^2 - 2\textrm{Re}\left(l_3l^*_0\bra{J_f}\hat{\mathcal{L}}_J(q)\ket{J_i}\bra{J_f}\hat{\mathcal{M}}_J(q)\ket{J_i}^*\right)\right] \Biggr\},	
\end{split} 
\end{equation}
where $q\equiv |\textbf{q}|$, and $\hat{\mathcal{M}}_J$, $\hat{\mathcal{L}}_J$, $\hat{\mathcal{T}}^{\textrm{el}}_J$, and $\hat{\mathcal{T}}^{\textrm{mag}}_J$ are the Coulomb, longitudinal, and transverse operators defined in terms of the hadron current $\mathcal{J}_{\mu}$. We suppress the magnetic quantum number $M$ of said operators throughout this paper, i.e., we denote, in the case of the Coulomb operator for example, $\hat{\mathcal{M}}_{JM}\equiv\hat{\mathcal{M}}_{J}$. These operators have both a vector (V) and an axial-vector (A) part, e.g. $\hat{\mathcal{M}}_J = \hat{M}_J^{\textrm{V}}-\hat{M}_J^{\textrm{A}}$, meaning that there are a total of eight operators. The operators are discussed in more detail in, e.g., \cite{Donnelly:1978tz, Walecka:1995mi, Ydrefors2011a}, and it is through them and their matrix elements that the chosen nuclear many-body framework enters into the scattering cross-section calculations. These matrix elements and their values are discussed in more detail in Sec.~\ref{sec:results} and the Appendix~\ref{sec:appendix}.

The double-differential scattering cross section to a final nuclear state with excitation energy $\omega = E_{K'} - E_{K} = E_{\nu} - E_{\nu'}$ can be shown to be \cite{Donnelly:1978tz, Walecka:1995mi, Ydrefors2011a}
\begin{equation}
\label{eq:xsec_costheta}
	\begin{split}
	&\frac{\textrm{d}^2\sigma_{i\rightarrow f}}{\textrm{d}\Omega\textrm{d}\omega} = \frac{G^2|\textbf{k}'|E_{\nu'}}{\pi(2J_i+1)}\left(\sum_{J\geq0}\sigma^J_{\textrm{CL}} + \sum_{J\geq1}\sigma^J_{\textrm{T}}\right),
	\end{split}
\end{equation}
which is written in terms of the Coulomb-longitudinal
\begin{equation}
	\begin{split}
	&\sigma^J_{\textrm{CL}} = (1+\cos\theta)|(J_f||\hat{\mathcal{M}}_J(q)||J_i)|^2 + \left(1 + \cos\theta - 2\frac{E_{\nu}E_{\nu'}}{q^2}\sin^2\theta\right)|(J_f||\hat{\mathcal{L}}_J(q)||J_i)|^2 \\& + \frac{E_{\nu}-E_{\nu'}}{q}(1+\cos\theta)2\textrm{Re}\left[(J_f||\hat{\mathcal{L}}_J(q)||J_i)(J_f||\hat{\mathcal{M}}_J(q)||J_i)^*\right]
	\end{split}
\end{equation}
and transverse
\begin{equation}
	\begin{split}
	&\sigma^J_{\textrm{T}} = \left(1-\cos\theta + \frac{E_{\nu}E_{\nu'}}{q^2}\sin^2\theta\right)\left[|(J_f||\hat{\mathcal{T}}^{\textrm{el}}_J(q)||J_i)|^2 + |(J_f||\hat{\mathcal{T}}^{\textrm{mag}}_J(q)||J_i)|^2\right] \\& -\frac{(E_{\nu}-E_{\nu'})}{q}(1-\cos\theta)2\textrm{Re}\left[(J_f||\hat{\mathcal{T}}^{\textrm{mag}}_J(q)||J_i)(J_f||\hat{\mathcal{T}}^{\textrm{el}}_J(q)||J_i)^*\right]
	\end{split}
\end{equation}
contributions to the cross section. Integrating over the angular coordinates $\Omega$ and summing over the individual final nuclear states yields the total cross section 
$\sigma(E_{\nu})$ as a function of the energy of the incoming neutrino. The origin of the incoming neutrino is taken into account in the form of a normalized energy distribution $\frac{\textrm{d}N}{\textrm{d}E_{\nu}}(E_{\nu})$, and the folded cross section $\braket{\sigma}$ is obtained by integrating $\frac{\textrm{d}N}{\textrm{d}E_{\nu}}(E_\nu)\sigma(E_{\nu})$ over all $E_{\nu}$.  The energy distributions for the different types of solar neutrinos used in this work are discussed in Sec.~\ref{sec:results}.

\subsection{The recoil-energy formalism}

One of the main purposes in this work is to perform nuclear-structure calculations that take into account detector-specific quantities such as nuclear recoil thresholds $T_\mathrm{thres}$. Although in principle the latter can be  taken into account in the cross section given in Eq.~(\ref{eq:xsec_costheta}), a clear disadvantage of the formalism presented in Sec.~\ref{sec:formalism_costeta} is the absence of a direct connection with experimental observables. Therefore, in this subsection we  devote an effort to express our results in terms of the nuclear recoil energy $T$ in order to have a clear link of the present calculations with experimentally measurable quantities.   

We  begin our discussion by expressing the three-momentum transfer in terms of the nuclear recoil energy. In that case, the kinematics of the process imply that 
\begin{equation}
    |\mathbf{q}|^2 = \left(E_\nu -   E_{\nu'}\right)^2 + 2 E_\nu E_{\nu'} (1-\cos \theta) = 2 M T + T^2 \, ,
\end{equation}
where $E_\nu-E_{\nu'}= \omega + T$, with $M$ being the nuclear mass and $\omega$ the nuclear excitation energy. By equating the two expressions above and working in the limit  $ M \gg E_\nu$, we get  
\begin{equation}
    T \approx  \frac{E_\nu (E_\nu - \omega) (1-\cos \theta) + \omega^2/2}{M} \, ,
    \label{eq:T_in_terms_of_costheta}
\end{equation}
in agreement with Ref.~\cite{Bednyakov:2018mjd}.
The minimum and maximum recoil-energy limits can be readily obtained by noting that the scattering angle is taking values in the range $-1 \leq \cos \theta \leq 1$, as
\begin{equation}
        T_\mathrm{min} = \frac{\omega^2}{2M} \, , \qquad
        T_\mathrm{max} = \frac{(2 E_\nu - \omega)^2}{2 M} \, .
        \label{eq:recoil_limits}
\end{equation}
Notice also that in the limit of elastic scattering, i.e., $\omega \to 0$, the usual CE$\nu$NS recoil-energy limits are recovered. 

In the next step our aim is to express the cross section given in Eq.~(\ref{eq:xsec_general}) in terms of the nuclear recoil energy. In that case, the following change of variables is appropriate:
$\frac{\textrm{d} \sigma}{\textrm{d} T} = \frac{d \sigma}{\textrm{d} \cos \theta} \left| \frac{\textrm{d} \cos \theta}{\textrm{d} T}\right |$, where the Jacobian can be immediately obtained using Eq.~(\ref{eq:T_in_terms_of_costheta}), and reads
\begin{equation}
    \left| \frac{\textrm{d} \cos \theta}{\textrm{d} T}\right | = \frac{M}{E_\nu (E_\nu - \omega)} \, .
\end{equation}
The differential cross section with respect to the nuclear recoil energy is then given by~\footnote{The cross section is further modified by a factor of $\frac{E_{K'}}{M}\left[1 + \frac{E_{\nu}}{M}\left(1 - \frac{E_{\nu'}}{|\textbf{k}'|}\cos\theta\right)\right]^{-1} \approx \left[1 + \frac{T -\omega^2 /(2 M) }{E_\nu - \omega} \right]^{-1}$ when the effect of the nuclear recoil $\textbf{K}' = \textbf{k}' - \textbf{k}$ on the phase space is taken into account \cite{Walecka:1995mi}. To a good approximation we have $E_{K'}/M \approx 1$, and for the neutrinos considered in this paper we also have $E_{\nu}, \omega \ll M$, so the first-order correction to the cross section arising from this phase-space distortion is vanishingly small and can be safely neglected.}
\begin{equation}
   \frac{\textrm{d} \sigma}{\textrm{d} T} =   \frac{\textrm{d} \sigma}{\textrm{d} \cos \theta}    \frac{M}{E_\nu (E_\nu - \omega)} \, .
\end{equation}
Analytical expressions for the relevant lepton traces appearing in Eq.~(\ref{eq:xsec_general}) can be derived, and they take the forms
\begin{equation}
  \sum_\mathrm{spins} l_0 l_0^* =  \frac{4 E_\nu^2-4 E_\nu (T+\omega )-2 M T+\omega  (2 T+\omega )}{2
   E_\nu (E_\nu-\omega )} \, ,
\end{equation}
\begin{equation}
  \sum_\mathrm{spins}  l_3 l_0^* =  \frac{(T+\omega ) \left(4 E_{\nu }^2-2 M T-4 E_{\nu } (T+\omega )+\omega  (2
   T+\omega )\right)}{2 \sqrt{2} E_{\nu } \left(E_{\nu }-\omega \right)
   \sqrt{M T}} \, ,
\end{equation}
\begin{equation}
  \sum_\mathrm{spins} l_3l_3^* =  \frac{(T+\omega )^2 \left(4 E_{\nu }^2-2 M T-4 E_{\nu } (T+\omega )+\omega 
   (2 T+\omega )\right)}{4 E_{\nu } M T \left(E_{\nu }-\omega \right)},
\end{equation}
\begin{equation}
  \sum_\mathrm{spins} \frac{1}{2}(\mathbf{l} \cdot \mathbf{l^*} - l_3 l_3^*) =  \frac{(2 M T-\omega  (2 T+\omega )) \left(4 E_{\nu }^2+2 M T-4 E_{\nu }
   (T+\omega )+\omega  (2 T+\omega )\right)}{8 E_{\nu } M T \left(E_{\nu
   }-\omega \right)},
\end{equation}
\begin{equation}
\sum_\mathrm{spins}   \frac{-i}{2}(\mathbf{l} \times \mathbf{l^*})_3 =  \frac{\left(2 E_{\nu } - \omega \right) (2 M T-\omega  (2 T+\omega ))}{2
   \sqrt{2} E_{\nu } \left(E_{\nu } - \omega\right) \sqrt{M T}}  \, ,
\end{equation}
under the approximations of $T\ll E_\nu$ and $T \ll M$. It is worth noticing that the above expressions in the limit $\omega\to 0$ are reduced to those obtained in Ref.~\cite{Hoferichter:2020osn} for the case of CE$\nu$NS. Note also that the positive sign appearing in the leading term of $l_3l_3^*$ given in Ref.~\cite{Hoferichter:2020osn} should be corrected by a minus sign.

A few comments are in order. By observing the latter expressions, interesting relations between the lepton traces can obtained offering insight into their relative contributions to the cross section. In particular, it holds that
\begin{equation}
\begin{aligned}
       l_3l_0^* = & \frac{T+\omega}{\sqrt{2 M T}} \,  l_0l_0^* \approx \left(\frac{T+\omega}{q} \right) \, l_0l_0^* \, ,\\
       l_3l_3^* = & \frac{(T+\omega)^2}{2 M T} \,  l_0l_0^* \approx \left(\frac{T+\omega}{q} \right)^2 \, l_0l_0^* \, .
\end{aligned}
\end{equation}
Evidently, the term $l_3l_0^*$ ($l_3l_3^*$) is suppressed (doubly suppressed) compared to $l_0l_0^*$ since for actual calculations it holds that  $T\ll \omega$ and $\omega/q \approx 10\%$ or less~\footnote{Indeed, $q$ is in the ballpark of 10--40~MeV, $\omega$ ranges   1.1--11.8~MeV for $^{205}$Tl  (1.3--19.2~MeV for $^{203}$Tl), while $T$ is of the order of a few~keV.}. Therefore, in the neutrino-nucleus scattering cross section given in Eq.~(\ref{eq:xsec_general}) one expects the corresponding  terms that are proportional to $l_3l_0^*$ and  $l_3l_3^*$ to have a minor contribution. For the sake of completeness, at this point it should be mentioned  that for very tiny recoil energies, i.e., $T \to T_\mathrm{min}$,  it holds that $\omega \approx q$, which implies that $l_3l_0^* \approx l_3l_3^* \approx l_0l_0^*$. However, the latter case is practically irrelevant in view of the typical recoil thresholds involved in 
neutrino-scattering experiments. On the other hand, the leptonic trace $(\mathbf{l} \cdot \mathbf{l^*} - l_3 l_3^*) $ can also be written in terms of $l_0l_0^*$ as
\begin{equation}
   (\mathbf{l} \cdot \mathbf{l^*} - l_3 l_3^*) =  \left(1-\frac{\omega  (2 T+\omega )}{2 M T}\right) \left(l_0l_0^* + \frac{2 M
   T}{E_{\nu } \left(E_{\nu }-\omega \right)}\right)  \, ,
\end{equation}
which further simplifies to 
\begin{equation}
       (\mathbf{l} \cdot \mathbf{l^*} - l_3 l_3^*) \approx  \left(1-\frac{\omega^2}{q^2}\right) \left(l_0l_0^* + \frac{q^2}{ E_{\nu } \left(E_{\nu }-\omega \right)}\right)  \, .
\end{equation}
For the  typical order  keV recoil energies, detectable in neutrino-scattering experiments, as explained previously, it holds that $1-(\omega/q)^2 \approx 1$, and hence $(\mathbf{l} \cdot \mathbf{l^*} - l_3 l_3^*)  \approx l_0l_0^* + \frac{q^2}{ E_{\nu } \left(E_{\nu }-\omega \right)}$, i.e., it is always larger than $l_0 l_0^*$, with the only exception being the case of extremely tiny recoil energies for which one has $(\mathbf{l} \cdot \mathbf{l^*} - l_3 l_3^*)  \ll l_0l_0^*$.

\section{Results}
\label{sec:results}

The performed shell-model calculations in the present work are optimized for the computation of inelastic neutrino-nucleus cross sections induced by solar neutrinos. To be concrete, the computed excitation spectra cover all the final states up to about $\omega \approx 11$~MeV for $^{205}$Tl and $\omega \approx 19$~MeV for $^{203}$Tl, which is more than adequate for even $hep$ and $^{8}$B neutrinos. Therefore, the following results are also applicable to inelastic neutrino-nucleus cross-section calculations for any neutrino source with similar energy range. Typical such examples include reactor neutrinos and geoneutrinos. Moreover, the present calculations can cover partly and may be relevant also to further neutrino sources such as diffuse supernova-neutrino background, supernova bursts, and primordial black holes.

\subsection{Results for cross sections in terms of the energy of the incoming neutrino}

\begin{figure}[t]
    \centering
    \includegraphics[width=0.48 \textwidth]{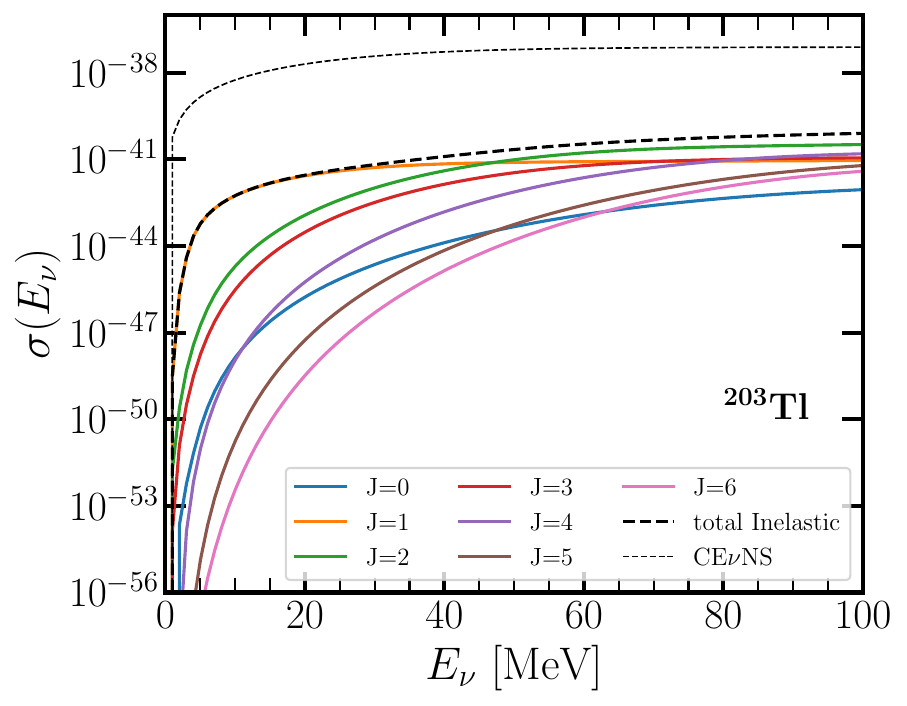}
    \includegraphics[width=0.48 \textwidth]{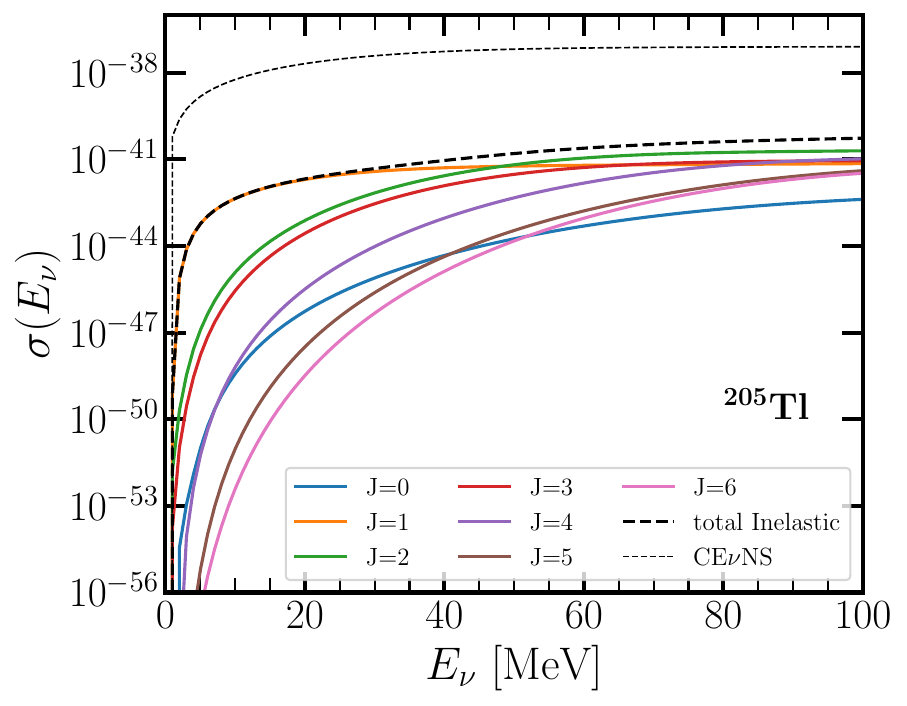}
    \caption{Integrated inelastic neutrino-nucleus cross sections as functions of the energy of the incident neutrino for $^{203}$Tl (left) and $^{205}$Tl (right). In both cases the results are given for the transitions to the final states of angular momentum $J$, adding up the $J^+$ and $J^-$ parity contributions. The corresponding CE$\nu$NS cross sections are also shown for comparison.}
    \label{fig:xsec_vs_Ev}
\end{figure}

We begin our discussion by presenting the inelastic neutrino-nucleus cross section as a function of the energy of the incoming neutrino in Fig.~\ref{fig:xsec_vs_Ev}, where the left and right graphs illustrate the results for the cases of $^{203}$Tl and $^{205}$Tl, respectively. The results are  demonstrated for the different $J$-transition contributions, neglecting recoil-energy thresholds. For both $^{203}$Tl and $^{205}$Tl isotopes, we expectedly find that the allowed $J=1^+$ transitions dominate the inelastic cross section for low neutrino energies (up to around $E_\nu \approx 43$~MeV), while for higher energies of the incoming neutrino the $J=2^+$ transitions have the dominant contribution (see the discussion below regarding the parities). For a review of the allowed approximation in inelastic neutrino scattering we refer the reader to \cite{Donnelly:1978tz, Walecka:1995mi}.   It is interesting to notice that in the high-energy regime, the $J=1,3,4,5,6$ transitions tend to have similar contributions to the inelastic cross section, with the remaining $J=2$ and $J=0$  having the largest and lowest contributions, respectively. For comparison purposes, the corresponding CE$\nu$NS cross sections are also given, from where we conclude that CE$\nu$NS dominates over the inelastic cross section by up to four orders of magnitude. Although not shown here, it is worth noting that we have performed a comparison of our inelastic cross section results with those coming out from the analytical formula presented in~\cite{Bednyakov:2018mjd}. In agreement with Ref.~\cite{Dutta:2022tav} we conclude that the formalism presented in~\cite{Bednyakov:2018mjd} leads to an overestimation of the cross section by up to four orders of magnitude, especially for neutrino energies that are much larger than the nuclear excitation energy.

To avoid overcrowding Fig.~\ref{fig:xsec_vs_Ev}, in the depicted results we have added the two parity contributions when plotting the inelastic cross sections for a given $J$-transition, i.e., we show the $J^+$ and $J^-$ contributions together.  Therefore, we wish to devote a separate paragraph for discussing the impact of the different parity contributions for a given $J$-transition. Specifically, the $J=0,1$ multipole transitions proceed only via positive-parity contributions for both nuclei of interest, since the final states with $J^{\pi}=1/2^-$ were omitted as explained in Sec.~\ref{sec:shell-model}. The $J=2^+$ multipole is dominant in comparison to $J=2^-$ in the full energy range  $0 \leq E_\nu \leq 100$~MeV of the incoming neutrino. Similarly, the $J=3^+$ multipole dominates over the $J=3^-$ multipole in the full range of neutrino energy. On the other hand, the $J=4^-$ multipole dominates over the $J=4^+$ one in the energy region $0 \leq E_\nu \leq 55 (50)$~MeV for the case of $^{203(205)}$Tl isotope, while for $E_\nu > 55 (50)$~MeV their behavior is reversed. A similar behavior is found regarding the $J=5$ multipole, i.e., the positive parity  dominates over the negative parity in the region $0 \leq E_\nu \leq 75 (55)$~MeV, while for higher neutrino energies the $J=5^-$ multipole becomes dominant. Finally, the $J=6^-$ multipole dominates over the $J=6^+$ in the full neutrino-energy range $0 \leq E_\nu \leq 100$~MeV, for both thallium isotopes. It is also interesting to note that the $J=5^-$ and $J=6^-$ multipoles have identical contributions to the inelastic cross section for both $^{203(205)}$Tl. For a visual illustration of these results, see Fig.~\ref{fig:xsec_vs_Ev_parities} in  the Appendix~\ref{sec:appendix}.

\begin{figure}[t!]
    \centering
    \includegraphics[width=0.48 \textwidth]{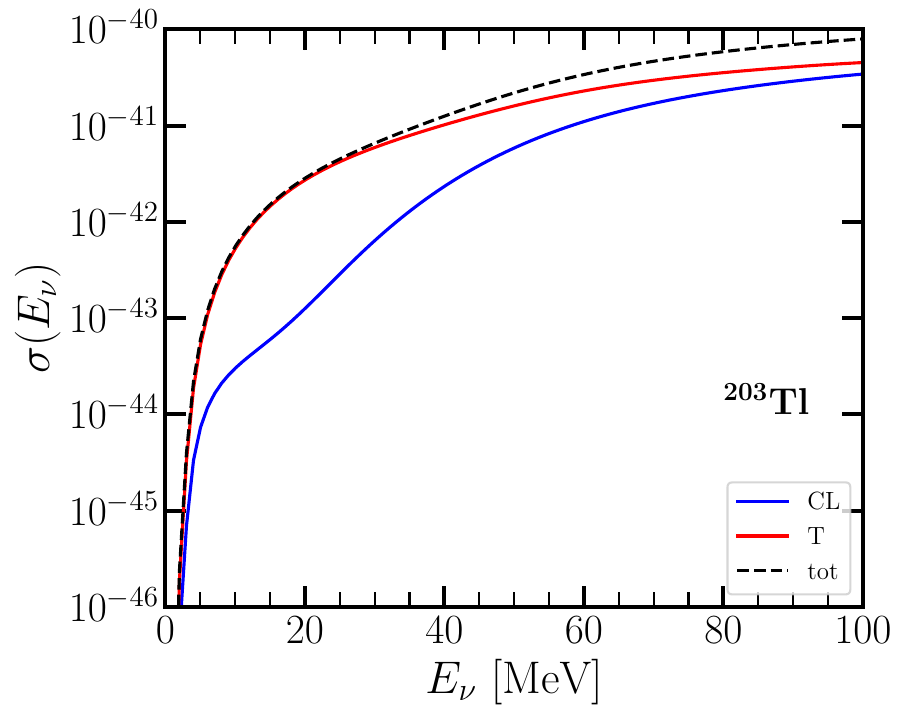}
    \includegraphics[width=0.48 \textwidth]{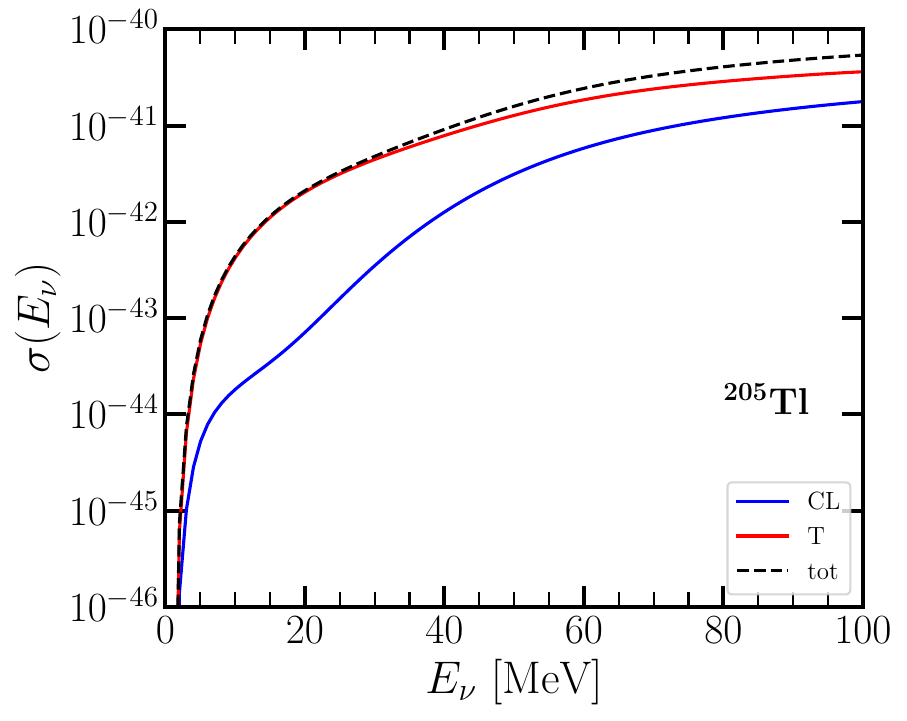}
    \caption{Same as Fig.~\ref{fig:xsec_vs_Ev}, but with the results given for the individual Coulomb-Longitudinal (CL) and Transverse (T) contributions to the inelastic cross section.}
    \label{fig:xsec_vs_Ev_operators}
\end{figure}

Next, in Fig.~\ref{fig:xsec_vs_Ev_operators} we present the relative contribution of the Coulomb-Longitudinal (CL) and transverse (T) operators to the total inelastic cross sections for $^{203}$Tl (left panel) and $^{205}$Tl (right panel). As previously, the results assume vanishing recoil threshold and are given in terms of the energy of the incident neutrino. Similar results are found for both isotopes. At very low energies, i.e., $0 \leq E_\nu \leq 5$~MeV, CL and T operators have similar contributions, whereas for $E_\nu \geq 5$~MeV the transverse operators dominate the total inelastic cross section, with CL being always subdominant, especially in the low and intermediate energy range $5 \leq E_\nu \leq 80$~MeV. Finally, it is interesting to notice that for higher neutrino energies, e.g. $E_\nu \geq 80$~MeV, CL and T contributions tend to be of similar size, especially in the case of $^{203}$Tl.

In Tables~\ref{tab:cross-sect-vs-energy-low} and \ref{tab:cross-sect-vs-energy-high} we present the individual contributions to the total inelastic cross sections, corresponding to pure vector ($\sigma_{\textrm{V}}$), axial-vector ($\sigma_{\textrm{A}}$), and interference ($\sigma_{\textrm{I}}$) components. For completeness, the total cross section ($\sigma_{\textrm{Tot.}}$) along with the Coulomb-longitudinal ($\sigma_{\textrm{CL}}$) and transverse ($\sigma_{\textrm{T}}$) contributions are also tabulated. To enable a refined analysis in the context of the different solar neutrinos, the two tables show the cross sections as functions of the neutrino energy in the ranges 0--2~MeV and 1--20~MeV, respectively.  It becomes evident that the axial-vector contribution dominates the inelastic cross sections, followed by the interference contribution, which is by about two orders of magnitude suppressed with respect to the former. The vector contribution is found to be further suppressed by about three to four (two) orders of magnitude compared to the axial-vector contribution in the range $0 \leq E_\nu \leq 10$~MeV ($E_\nu \geq 10$~MeV). 

\begin{table*}%[h]
\caption{Total inelastic neutral-current scattering cross sections $\sigma_{\textrm{Tot.}}$ off $^{203}$Tl and $^{205}$Tl as functions of the energy $E_{\nu}$ of the incoming neutrino. The contributions to the total cross section from the vector $\sigma_{\textrm{V}}$, axial-vector $\sigma_{\textrm{A}}$, and interference $\sigma_{\textrm{I}}$ parts along with the Coulomb-longitudinal $\sigma_{\textrm{CL}}$ and transverse $\sigma_{\textrm{T}}$ parts are also given. The format in which the data is presented is $R$($\textrm{e}$), and the cross sections are obtained by $\sigma(E_{\nu}) = R\times 10^{\textrm{e}}\times D$, the units being $D=10^{-50}$ cm$^2$.}
\resizebox{\columnwidth}{!}{
\begin{tabular}{ l|cccccc|cccccc } 
 \hline\hline
 \multirow{2}{*}{}&\multicolumn{12}{c}{Nucleus}\\
 \cline{2-7}\cline{8-13}&\multicolumn{6}{c|}{$^{203}$Tl} & \multicolumn{6}{c}{$^{205}$Tl} \\
 %\cline{2-5}\cline{6-9}\\
 \cline{2-7}\cline{8-13}$E_{\nu}$ (MeV)& $\sigma_{\textrm{V}}$ & $\sigma_{\textrm{A}}$ & $\sigma_{\textrm{I}}$ & $\sigma_{\textrm{CL}}$ & $\sigma_{\textrm{T}}$ & $\sigma_{\textrm{Tot.}}$ & $\sigma_{\textrm{V}}$ & $\sigma_{\textrm{A}}$ & $\sigma_{\textrm{I}}$ & $\sigma_{\textrm{CL}}$ & $\sigma_{\textrm{T}}$ & $\sigma_{\textrm{Tot.}}$\\ \hline
  0.1&0(0)&0(0)&0(0)&0(0)&0(0)&0(0)&0(0)&0(0)&0(0)&0(0)&0(0)&0(0)\\
  0.2&0(0)&0(0)&0(0)&0(0)&0(0)&0(0)&2.024(-8)&3.227(-3)&-1.348(-5)&9.661(-4)&2.248(-3)&3.214(-3)\\
  0.3&2.256(-6)&4.226(-1)&1.399(-3)&8.198(-2)&3.420(-1)&4.240(-1)&3.417(-6)&1.526(-1)&-1.226(-3)&2.595(-2)&1.255(-1)&1.514(-1)\\
  0.4&2.399(-5)&1.802(0)&8.760(-3)&2.259(-1)&1.585(0)&1.811(0)&2.802(-5)&5.238(-1)&-6.230(-3)&5.876(-2)&4.588(-1)&5.176(-1)\\
  0.5&1.093(-4)&4.145(0)&2.654(-2)&3.651(-1)&3.806(0)&4.171(0)&1.187(-4)&1.114(0)&-1.757(-2)&9.046(-2)&1.006(0)&1.097(0)\\
  0.6&3.522(-4)&7.458(0)&5.922(-2)&4.883(-1)&7.029(0)&7.517(0)&3.676(-4)&1.920(0)&-3.776(-2)&1.201(-1)&1.762(0)&1.883(0)\\
  0.7&9.339(-4)&1.175(1)&1.113(-1)&5.950(-1)&1.127(1)&1.186(1)&9.424(-4)&2.936(0)&-6.926(-2)&1.483(-1)&2.720(0)&2.868(0)\\
  0.8&2.173(-3)&1.704(1)&1.876(-1)&6.873(-1)&1.654(1)&1.723(1)&2.126(-3)&4.158(0)&-1.145(-1)&1.757(-1)&3.870(0)&4.045(0)\\
  0.9&4.605(-3)&2.333(1)&2.925(-1)&7.673(-1)&2.286(1)&2.363(1)&4.392(-3)&5.577(0)&-1.757(-1)&2.034(-1)&5.202(0)&5.406(0)\\
  1.0&9.160(-3)&3.065(1)&4.312(-1)&8.379(-1)&3.025(1)&3.109(1)&8.531(-3)&7.187(0)&-2.552(-1)&2.325(-1)&6.708(0)&6.941(0)\\
  1.1&1.740(-2)&3.960(1)&6.063(-1)&1.086(0)&3.913(1)&4.022(1)&1.581(-2)&8.980(0)&-3.550(-1)&2.646(-1)&8.376(0)&8.641(0)\\
  1.2&3.180(-2)&5.399(1)&8.072(-1)&2.523(0)&5.231(1)&5.483(1)&2.868(-2)&1.531(2)&-4.957(-2)&4.338(1)&1.097(2)&1.531(2)\\
  1.3&5.620(-2)&9.595(1)&1.083(0)&1.151(1)&8.558(1)&9.709(1)&5.284(-2)&7.132(2)&2.106(0)&1.914(2)&5.239(2)&7.153(2)\\
  1.4&9.649(-2)&2.118(2)&1.638(0)&3.904(1)&1.745(2)&2.136(2)&9.566(-2)&2.228(3)&7.742(0)&5.792(2)&1.656(3)&2.235(3)\\
  1.5&1.622(-1)&6.935(2)&3.337(0)&1.701(2)&5.269(2)&6.970(2)&1.693(-1)&6.050(3)&2.075(1)&1.525(3)&4.546(3)&6.071(3)\\
  1.6&2.668(-1)&2.142(3)&7.908(0)&5.504(2)&1.600(3)&2.150(3)&2.887(-1)&1.232(4)&4.377(1)&2.917(3)&9.442(3)&1.236(4)\\
  1.7&4.278(-1)&4.694(3)&1.676(1)&1.159(3)&3.552(3)&4.711(3)&4.725(-1)&2.108(4)&7.929(1)&4.654(3)&1.650(4)&2.116(4)\\
  1.8&6.682(-1)&8.562(3)&3.136(1)&2.005(3)&6.589(3)&8.594(3)&7.435(-1)&3.245(4)&1.299(2)&6.682(3)&2.590(4)&3.258(4)\\
  1.9&1.017(0)&1.408(4)&5.378(1)&3.134(3)&1.101(4)&1.414(4)&1.133(0)&4.833(4)&2.029(2)&9.502(3)&3.903(4)&4.853(4)\\
  2.0&1.515(0)&2.171(4)&8.751(1)&4.618(3)&1.718(4)&2.180(4)&1.679(0)&6.963(4)&3.045(2)&1.321(4)&5.673(4)&6.994(4)\\
 \hline
 \hline
\end{tabular}
}
\label{tab:cross-sect-vs-energy-low}
\end{table*}
\begin{table*}%[h]
\caption{Same as table \ref{tab:cross-sect-vs-energy-low}, but for higher energies of the incoming neutrino and in units of $D = 10^{-43}$ cm$^2$.}
\resizebox{\columnwidth}{!}{
\begin{tabular}{ l|cccccc|cccccc } 
 \hline\hline
 \multirow{2}{*}{}&\multicolumn{12}{c}{Nucleus}\\
 \cline{2-7}\cline{8-13}&\multicolumn{6}{c|}{$^{203}$Tl} & \multicolumn{6}{c}{$^{205}$Tl} \\
 %\cline{2-5}\cline{6-9}\\
 \cline{2-7}\cline{8-13}$E_{\nu}$ (MeV)& $\sigma_{\textrm{V}}$ & $\sigma_{\textrm{A}}$ & $\sigma_{\textrm{I}}$ & $\sigma_{\textrm{CL}}$ & $\sigma_{\textrm{T}}$ & $\sigma_{\textrm{Tot.}}$ & $\sigma_{\textrm{V}}$ & $\sigma_{\textrm{A}}$ & $\sigma_{\textrm{I}}$ & $\sigma_{\textrm{CL}}$ & $\sigma_{\textrm{T}}$ & $\sigma_{\textrm{Tot.}}$\\ \hline
  1.0&9.160(-10)&3.065(-6)&4.312(-8)&8.379(-8)&3.025(-6)&3.109(-6)&8.531(-10)&7.187(-7)&-2.552(-8)&2.325(-8)&6.708(-7)&6.941(-7)\\
  2.0&1.515(-7)&2.171(-3)&8.751(-6)&4.618(-4)&1.718(-3)&2.180(-3)&1.679(-7)&6.963(-3)&3.045(-5)&1.321(-3)&5.673(-3)&6.994(-3)\\
  3.0&3.196(-6)&3.691(-2)&2.208(-4)&6.493(-3)&3.064(-2)&3.713(-2)&3.052(-6)&7.179(-2)&4.574(-4)&9.978(-3)&6.227(-2)&7.225(-2)\\
  4.0&2.482(-5)&2.081(-1)&1.657(-3)&3.201(-2)&1.778(-1)&2.098(-1)&2.047(-5)&2.510(-1)&2.221(-3)&2.764(-2)&2.256(-1)&2.532(-1)\\
  5.0&1.122(-4)&5.753(-1)&5.973(-3)&7.133(-2)&5.101(-1)&5.814(-1)&8.389(-5)&5.712(-1)&6.547(-3)&5.120(-2)&5.266(-1)&5.778(-1)\\
  6.0&3.679(-4)&1.147(0)&1.483(-2)&1.157(-1)&1.046(0)&1.162(0)&2.577(-4)&1.038(0)&1.468(-2)&7.700(-2)&9.755(-1)&1.052(0)\\
  7.0&9.805(-4)&1.920(0)&2.983(-2)&1.613(-1)&1.789(0)&1.951(0)&6.556(-4)&1.647(0)&2.778(-2)&1.029(-1)&1.572(0)&1.675(0)\\
  8.0&2.262(-3)&2.890(0)&5.246(-2)&2.070(-1)&2.738(0)&2.945(0)&1.461(-3)&2.395(0)&4.695(-2)&1.285(-1)&2.315(0)&2.443(0)\\
  9.0&4.687(-3)&4.049(0)&8.412(-2)&2.527(-1)&3.886(0)&4.138(0)&2.950(-3)&3.276(0)&7.319(-2)&1.538(-1)&3.198(0)&3.352(0)\\
  10.0&8.945(-3)&5.390(0)&1.261(-1)&2.989(-1)&5.226(0)&5.525(0)&5.515(-3)&4.283(0)&1.074(-1)&1.792(-1)&4.217(0)&4.396(0)\\
  11.0&1.598(-2)&6.903(0)&1.794(-1)&3.468(-1)&6.751(0)&7.098(0)&9.692(-3)&5.411(0)&1.504(-1)&2.052(-1)&5.365(0)&5.571(0)\\
  12.0&2.706(-2)&8.577(0)&2.452(-1)&3.976(-1)&8.451(0)&8.849(0)&1.619(-2)&6.650(0)&2.029(-1)&2.328(-1)&6.636(0)&6.869(0)\\
  13.0&4.379(-2)&1.040(1)&3.242(-1)&4.531(-1)&1.032(1)&1.077(1)&2.591(-2)&7.994(0)&2.655(-1)&2.629(-1)&8.023(0)&8.286(0)\\
  14.0&6.820(-2)&1.237(1)&4.172(-1)&5.156(-1)&1.234(1)&1.285(1)&3.999(-2)&9.436(0)&3.388(-1)&2.967(-1)&9.518(0)&9.815(0)\\
  15.0&1.028(-1)&1.446(1)&5.247(-1)&5.877(-1)&1.450(1)&1.509(1)&5.979(-2)&1.097(1)&4.231(-1)&3.356(-1)&1.111(1)&1.145(1)\\
  16.0&1.505(-1)&1.667(1)&6.472(-1)&6.726(-1)&1.680(1)&1.747(1)&8.697(-2)&1.258(1)&5.188(-1)&3.815(-1)&1.281(1)&1.319(1)\\
  17.0&2.149(-1)&1.900(1)&7.851(-1)&7.739(-1)&1.922(1)&2.000(1)&1.234(-1)&1.427(1)&6.264(-1)&4.363(-1)&1.459(1)&1.502(1)\\
  18.0&3.000(-1)&2.142(1)&9.388(-1)&8.959(-1)&2.176(1)&2.266(1)&1.715(-1)&1.603(1)&7.460(-1)&5.024(-1)&1.645(1)&1.695(1)\\
  19.0&4.105(-1)&2.393(1)&1.109(0)&1.043(0)&2.440(1)&2.545(1)&2.335(-1)&1.786(1)&8.781(-1)&5.822(-1)&1.839(1)&1.897(1)\\
  20.0&5.515(-1)&2.652(1)&1.295(0)&1.221(0)&2.715(1)&2.837(1)&3.126(-1)&1.975(1)&1.023(0)&6.787(-1)&2.041(1)&2.108(1)\\
 \hline
 \hline
\end{tabular}
}
\label{tab:cross-sect-vs-energy-high}
\end{table*}

\subsection{Cross sections in terms of the nuclear recoil}

\begin{figure}[t!]
    \centering
    \includegraphics[width=0.48 \textwidth]{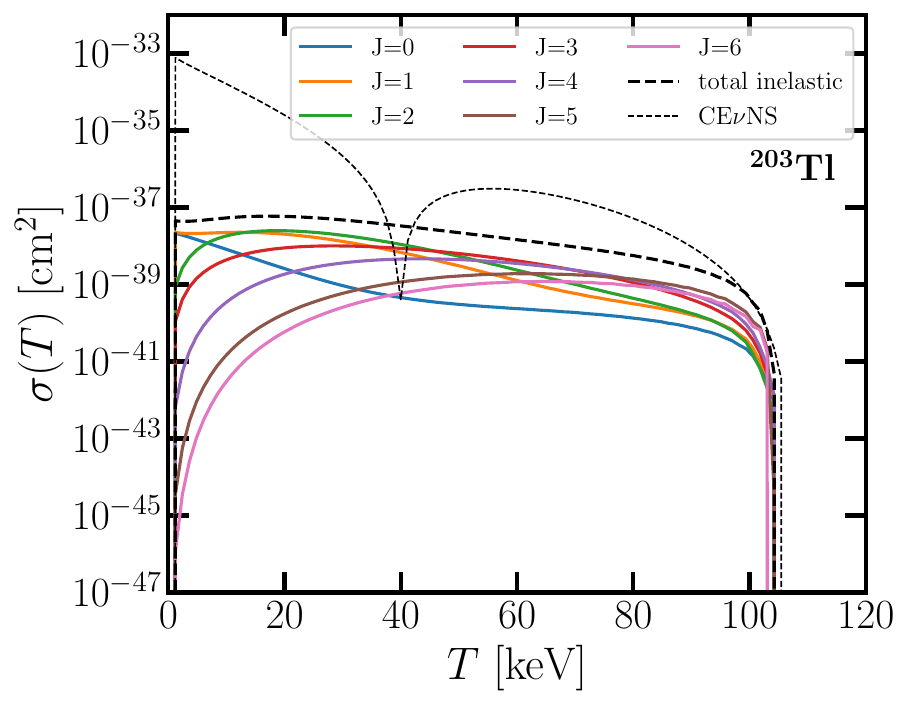}
    \includegraphics[width=0.48 \textwidth]{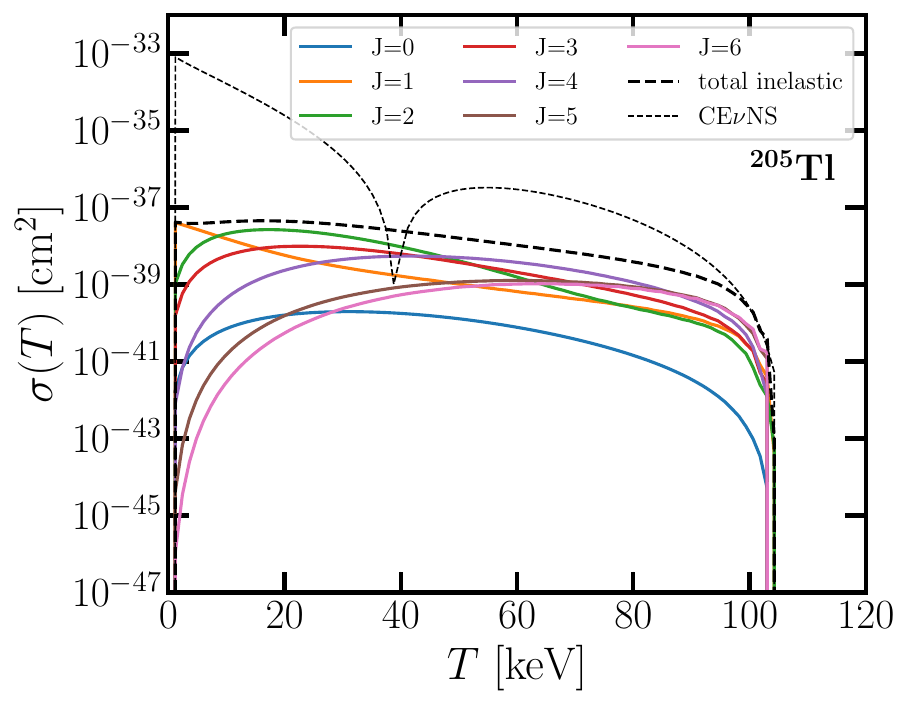}
    \caption{Same as Fig.~\ref{fig:xsec_vs_Ev}, but with the results given in terms of the nuclear recoil energy.}
    \label{fig:xsec_vs_reco}
\end{figure}

We now turn our discussion to the inelastic cross sections calculated by taking into account also the nuclear recoil energy, $T$. After integrating over the energies of the incoming neutrino, we obtain the cross sections as functions of the nuclear recoil energy, as depicted in Fig.~\ref{fig:xsec_vs_reco}. As previously, the left and right panels correspond to $^{203}$Tl and $^{205}$Tl, while the individual contributions for the various $J$-transitions are also given. For both cases the $J=1$ transition dominates the cross section for very low recoil energies, while above $T \approx 10$~keV the $J=2$ contribution becomes the most relevant one.  The remaining transitions behave similarly for both isotopes, with the only exception being the $J=0$ one, which is relevant for very low recoil energies in the case of $^{203}$Tl only. We furthermore superimpose the corresponding CE$\nu$NS-integrated cross section for the sake of comparison. The latter, as expected, dominates over the inelastic channels. However, it is interesting to notice the dip occurring at about $T=40$~keV, which reflects the loss of coherence that is taken into account via the 
ground-state nuclear form factor. In this particular region the inelastic channel becomes dominant and precise neutrino-nucleus cross-sections measurements could shed light on such subtle occurrences of coherence and incoherence. 

\subsection{Folded solar-neutrino cross sections}

To take into account the origin of the incoming neutrino, we have folded our computed neutrino-nucleus cross sections, given as functions of the neutrino energy, over energy distributions corresponding to different types of solar neutrinos. The distributions of $pep$, and $^{7}$Be neutrinos are monochromatic with energies of 1.442 MeV and 862 keV, respectively. The $e^- + \prescript{7}{}{\textrm{Be}} \rightarrow \nu_{e} + \prescript{7}{}{\textrm{Li}}$ reaction that produces $^{7}$Be neutrinos can also leave the $^{7}$Li nucleus in its first excited state, in which case the energy of the produced neutrino will be 384 keV. We will denote the higher-energy $^{7}$Be neutrinos as $^7\mathrm{Be(high)}$ and the 
lower-energy neutrinos as $^7\mathrm{Be(low)}$.

The energy distributions of the other solar neutrinos are continuous. For the $pp$, $^{13}$N, $^{15}$O, and $^{17}$F neutrinos we have utilized the spectra~\cite{Raffelt:1996wa}
\begin{equation}
    \frac{\textrm{d}N}{\textrm{d}E_{\nu}}(E_{\nu}) = N_0(Q + m_{e} - E_{\nu})E^2_{\nu}\sqrt{(Q + m_{e} - E_{\nu})^2 - m_{e}^{2}},
\end{equation}
where $Q$ is the $Q$-value of the weak reaction producing the neutrino and $N_0$ is a normalization constant which guarantees that
\begin{equation}
    \int_0^{\infty}\textrm{d}E_{\nu}\frac{\textrm{d}N}{\textrm{d}E_{\nu}}(E_{\nu}) = 1.
\end{equation}
For $^{8}$B and $hep$ we have similarly used
\begin{equation}
    \frac{\textrm{d}N}{\textrm{d}E_{\nu}}(E_{\nu}) = N_0E^2_{\nu}(Q - E_{\nu})^{11/4}
\end{equation}
and
\begin{equation}
    \frac{\textrm{d}N}{\textrm{d}E_{\nu}}(E_{\nu}) = N_0E^{48/25}_{\nu}(Q - E_{\nu})^{9/5},
\end{equation}
respectively.  The $Q$-values of the reactions that produce neutrinos with continuous spectra are  tabulated in Table~\ref{tab:Q-values}. To relate the reaction cross sections obtained by folding $\sigma(E_{\nu})$ over the above neutrino-energy distributions to experimentally observable scattering cross sections, we must take into account the fluxes of different types of solar neutrinos, which we do in the form of appropriate flux normalization factors \cite{Baxter:2021pqo}.
\begin{table}%[h]
\caption{The $Q$-values (MeV) of the reactions that produce solar neutrinos with continuous energy spectra.}
\begin{tabular}{ lcr } 
 \hline
 \hline
 Neutrino & Reaction & $Q$-value \\ \hline 
 $pp$&$p + p \rightarrow \nu_{e} + \prescript{2}{}{\textrm{H}} + e^+$&0.420\\
 $^{13}$N&$\prescript{13}{}{\textrm{N}} \rightarrow \nu_{e} + \prescript{13}{}{\textrm{C}} + e^+$&1.199\\
 $^{15}$O&$\prescript{15}{}{\textrm{O}} \rightarrow \nu_{e} + \prescript{15}{}{\textrm{N}} + e^+$&1.732\\
 $^{17}$F&$\prescript{17}{}{\textrm{F}} \rightarrow \nu_{e} + \prescript{17}{}{\textrm{O}} + e^+$&1.740\\
 $^{8}$B&$\prescript{8}{}{\textrm{B}} \rightarrow \nu_{e} + \prescript{8}{}{\textrm{Be}}^* + e^+$&15.1\\
 $hep$&$\prescript{3}{}{\textrm{He}} + p \rightarrow \nu_{e} + \prescript{4}{}{\textrm{He}} + e^+$&18.77\\ \hline
 \hline
\end{tabular}
\label{tab:Q-values}
\end{table}

\begin{figure}[t!]
    \centering
    \includegraphics[width=0.49 \textwidth]{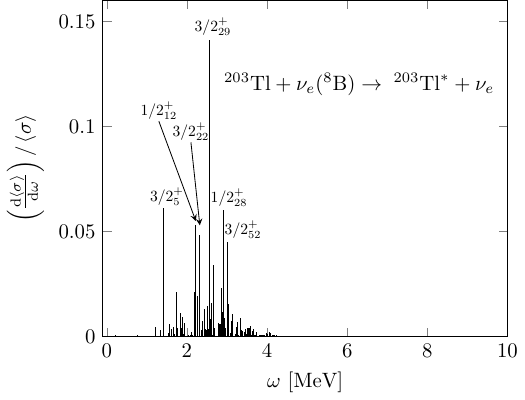}
    \includegraphics[width=0.49 \textwidth]{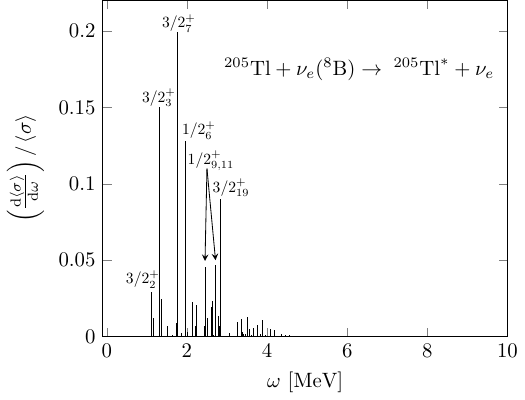}
    \caption{Contributions from individual nuclear final states, with excitation energy $\omega$, to the total folded $^{8}$B solar-neutrino scattering cross sections for both nuclei under discussion. Individual contributions are normalized to the total folded cross section and thus represent the fraction which scattering to that particular nuclear final state contributes to the total folded cross section. }
    \label{fig:contributions-final-states-8B}
\end{figure}

\begin{figure}[t!]
    \centering
    \includegraphics[width=0.49 \textwidth]{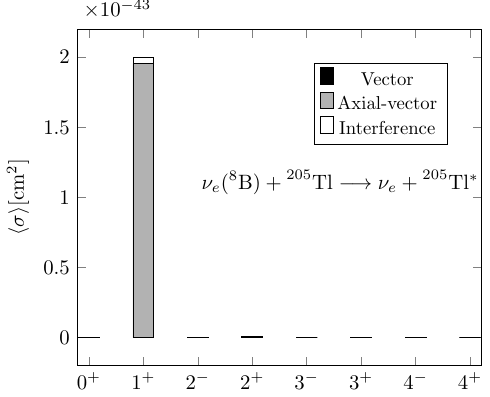}
    \includegraphics[width=0.49 \textwidth]{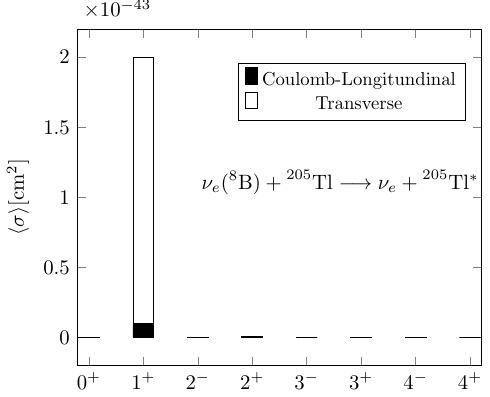}
    \caption{Reaction cross-section contributions for $^{8}$B neutrino scattering off $^{205}$Tl, with the neutrino flux normalized to 1, from different transition multipoles decomposed into vector, axial-vector, and interference parts (left), and into Coulomb-longitudinal and transverse parts (right).}
    \label{fig:multipoles}
\end{figure}

The contributions to the total $^{8}$B solar-neutrino scattering cross section from individual final nuclear states are illustrated in Fig.~\ref{fig:contributions-final-states-8B}. These contribution profiles are quite similar for both considered nuclei. Some of the most strongly contributing states are labeled in the figure, and all of them are reached from the ground states ($J^{\pi}_{\textrm{g.s.}} = 1/2^+_1$ for both nuclei) through an allowed transition. As discussed previously, the forbidden transitions have an almost negligible contribution for such low-energy neutrino sources. This is further demonstrated in Fig.~\ref{fig:multipoles} for $^{205}$Tl. Due to the low energy of the neutrinos, the contributions from individual final states come mostly from a relatively small number of states scattered in energy in the range $\approx 1-4$ MeV. This is in contrast with higher-energy astrophysical neutrinos such as supernova neutrinos, for example, for which evidence exists that indicates that the neutral-current scattering cross section comes mostly from spin-flip M1 giant resonances \cite{Hellgren:2022yvo}. 

\begin{figure}[t!]
    \centering
    \includegraphics[width=0.48 \textwidth]{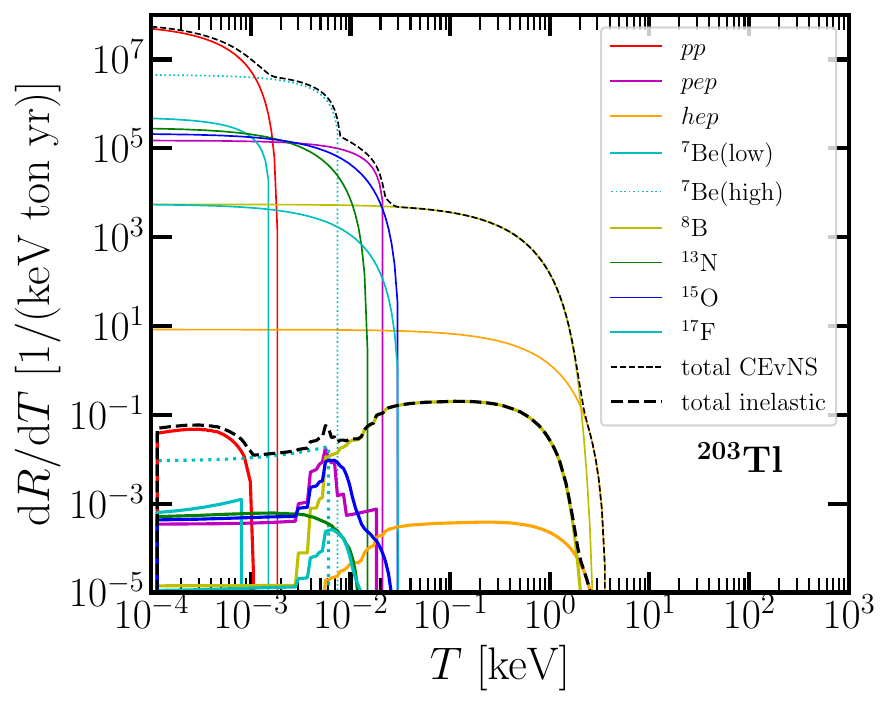}
    \includegraphics[width=0.48 \textwidth]{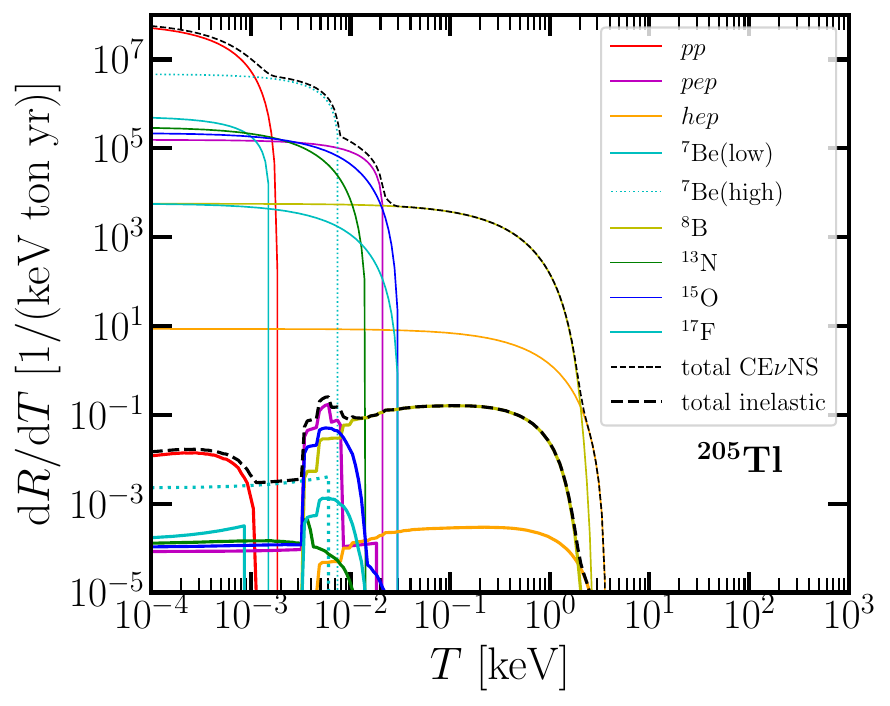}
    \caption{Differential number of events as a function of the nuclear recoil energy for the case of $^{203}$Tl (left) and $^{205}$Tl (right). Individual spectra are shown for the different solar-neutrino sources. Thick (thin) curves correspond to the inelastic (CE$\nu$NS) rates.}
    \label{fig:dR_vs_reco}
\end{figure}

Next, in Fig.~\ref{fig:dR_vs_reco}, we evaluate the differential number of events due to inelastic neutrino-nucleus scattering off  $^{203}$Tl (left panel) and $^{205}$Tl (right panel) induced by solar neutrinos, as
\begin{equation}
    \frac{\mathrm{d} R_i}{\mathrm{d} T} = \mathcal{E} \sum_\omega \int_{E_\nu^\mathrm{min}(T, \omega)}^ {E_\nu^\mathrm{max}} 
 \frac{\mathrm{d} N_i}{\mathrm{d} E_\nu} (E_\nu) \frac{\mathrm{d} \sigma}{\mathrm{d} T} (E_\nu, T, \omega) \, \mathrm{d} E_\nu \, ,
\end{equation}
where $i$ refers to the various solar-neutrino sources, and the exposure is set at $\mathcal{E}=1~\mathrm{ton ~ yr}$. The upper integration limit is obtained from the 
endpoint of the $i$-th solar-neutrino source, while the lower one is taken by inverting the expression in the right-hand side of Eq.~(\ref{eq:recoil_limits}).
The results are illustrated in terms of the nuclear recoil energy, while individual rates are given for the various solar-neutrino sources. As expected, the induced recoil signal is very tiny and up to a few keV. Let us also stress that in view of the current detector technology a nuclear recoil signal below 0.1~keV is challenging to be achieved, while most dark matter direct detection detectors are sensitive in the region of a few keV and above. Hence, the most relevant spectra are those induced by $^8$B neutrinos. Compared to the corresponding CE$\nu$NS rates which are also shown here and highlighted with the same color code (thin curves), one concludes that the inelastic scattering channel leads to a suppressed signal. It is interesting to note that unlike the CE$\nu$NS case, the individual signal induced by the $^8$B neutrino flux is the most pronounced among the various inelastic rates. This is because of the numerous excited nuclear states that lie within the energy range of $^8$B neutrinos. On the other hand, for the case of CE$\nu$NS the rates are mostly driven by the neutrino-flux normalizations and much less by the nuclear-physics aspects. Indeed, recalling Fig.~\ref{fig:xsec_vs_reco}, it can be seen that for nuclear-physics effects to become relevant in the CE$\nu$NS rates, a neutrino source of higher energy, such as atmospheric neutrinos, is required to trigger recoil energies in the ballpark of few tens of keV.

\begin{figure}[t!]
    \centering
    \includegraphics[width=0.48 \textwidth]{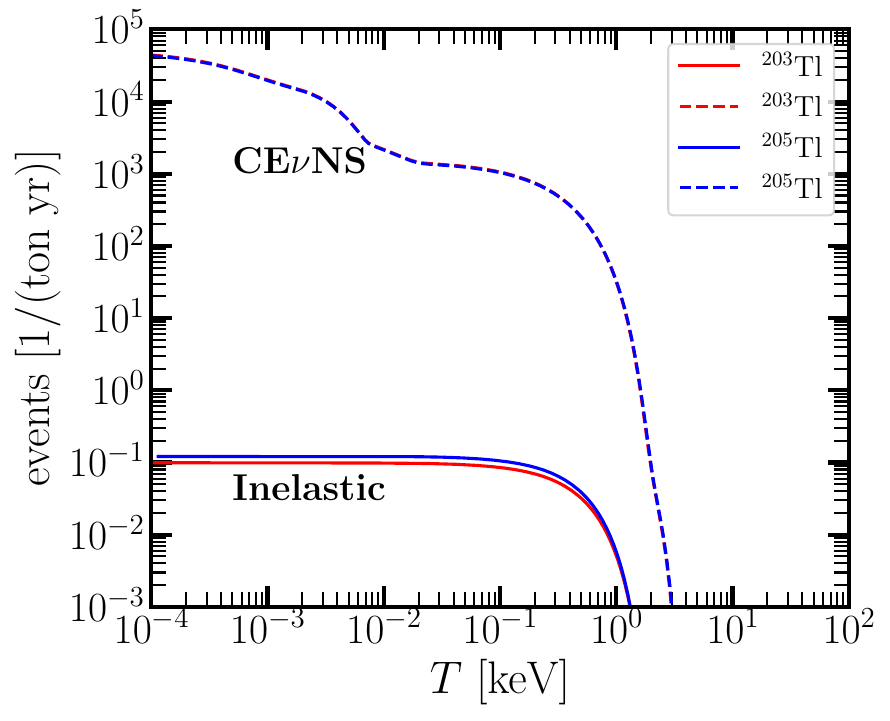}
    \caption{Total number of events induced by solar neutrino-nucleus scattering  as a function of the nuclear recoil energy for the cases of $^{203}$Tl (red color)  and $^{205}$Tl (blue color). The inelastic (CE$\nu$NS) rates are shown with solid (dashed) curves.}
    \label{fig:Events_vs_reco}
\end{figure}

In Fig.~\ref{fig:Events_vs_reco}, the integrated event rates above threshold are depicted. In the nuclear recoil region of interest the CE$\nu$NS signal dominates by four orders of magnitude. An interesting feature concerning the inelastic rates is that the signal remains constant for a low recoil threshold, and hence there is no need to achieve even lower recoil thresholds for getting enhanced rates as in the case of CE$\nu$NS. As already explained previously, comparable rates are expected for higher recoil energies which, in turn, require more energetic neutrino sources such as pion-decay-at-rest, diffuse supernova, and atmospheric neutrinos. The latter will also lead to recoil features in the regime of a few tens of keV where the corresponding CE$\nu$NS signal suffers from loss of coherence. Since the present shell-model calculations are not optimized to describe highly excited nuclear final states, this exercise is left for a future study.  Before closing this discussion we would like to stress, however, that although the present inelastic rates are quite suppressed, they may be comparable to new-physics effects which are traditionally probed using only the CE$\nu$NS channel. Hence for a more accurate sensitivity extraction the present calculations are relevant.

\section{Conclusions}
\label{sec:conclusions}

In this work we  focused on low-energy neutral-current neutrino-nucleus scattering off stable thallium isotopes. Specifically, our computed results involve inelastic cross sections and event rates of solar neutrinos scattering off  $^{203,205}$Tl. From the experimental point of view, these isotopes are of particular interest, being the dopant material in several dark-matter direct-detection experiments such as COSINE, DAMA/LIBRA, PICO-LON, ANAIS, and SABRE. Based on extensive shell-model calculations, we  performed a thorough study of all accessible nuclear final states covering the energy range of solar neutrinos. The inelastic neutrino-$^{203,205}$Tl cross sections as well as the corresponding event rates were found to be dominated by axial-vector interactions. Concerning  operators, the transverse ones  dominate the cross section across all energies in the range $0-100$~MeV. The most important transitions are determined to be of multipolarity $J=1^+$ ($J=2^+$) in the low (high) energy regime. For solar neutrinos, in particular, the allowed $J=1^+$ transition is the most relevant one.

In our effort to achieve a direct connection between our present calculations and experimental observables, our computed cross sections and corresponding event rates take into account the nuclear recoil energy, i.e., a key quantity that is traditionally ignored in previous similar studies. For the first time, we  expressed the inelastic neutrino-nucleus cross-section formalism in terms of the nuclear recoil energy instead of the scattering angle. We  furthermore focused our attention on the various lepton traces that are proportional to the respective matrix elements entering the generic inelastic neutrino-nucleus cross-section formula. By expressing these quantities in terms of the energy of the incoming neutrino,  nuclear excitation energy and nuclear recoil energy, we ended up with insightful relations between them. This allowed us to draw conclusions regarding the relative contribution of each term in the inelastic neutrino-nucleus cross section. 

We  finally discussed how the inelastic neutrino-nucleus cross sections explored here compare to the dominant CE$\nu$NS channel. Although the CE$\nu$NS-induced solar-neutrino rates are found to be by up to four orders of magnitude larger, we  nonetheless stressed that the inelastic channel may be relevant in analyses involving scenarios beyond the standard model physics.   We  also remarked that the inelastic rates can be comparable or even exceed the CE$\nu$NS ones. For the thallium isotopes studied here, this corresponds to a recoil energy of about 40~keV, i.e., a region of the momentum transfer around  the first dip of the vector ground-state-to-ground-state nuclear form factor where   a sharp loss of coherence is occurring. This will be more relevant for more energetic neutrino sources, such as pion-decay-at-rest, supernova, and atmospheric neutrinos.

\section*{Acknowledgments}

DKP wishes to thank Pablo Mu\~noz Candela for fruitful discussions.
The work of DKP was supported by the Hellenic Foundation for Research and Innovation (H.F.R.I.) under the “3rd Call for H.F.R.I. Research Projects to support Post-Doctoral Researchers” (Project Number: 7036). MH acknowledges financial support from the V\"ais\"al\"a Foundation of the Finnish Academy of Science and Letters.

\appendix

\section{}
\label{sec:appendix}

\begin{figure}[t]
    \centering
    \includegraphics[width=0.48 \textwidth]{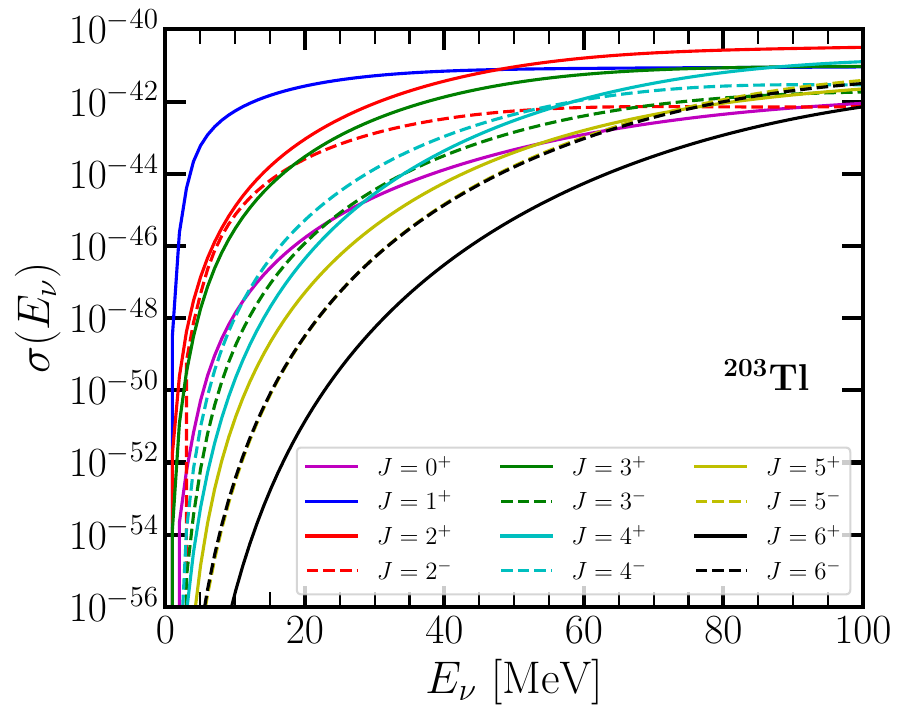}
    \includegraphics[width=0.48 \textwidth]{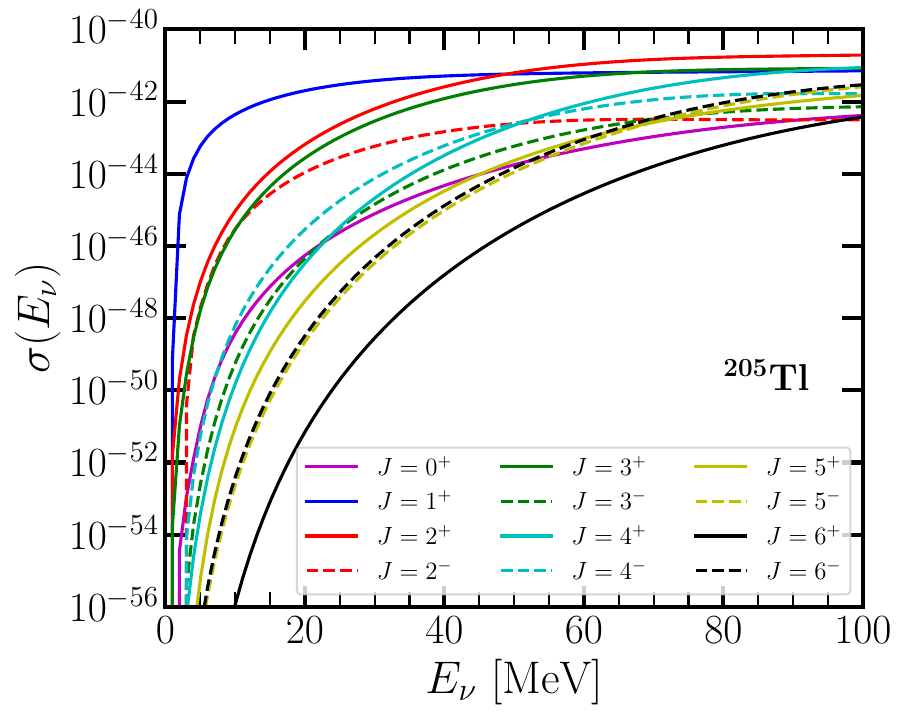}
    \caption{Same as Fig.~\ref{fig:xsec_vs_Ev}, but with the results given separately for the different $J^\pm$-transitions.}
    \label{fig:xsec_vs_Ev_parities}
\end{figure}

In this appendix we reproduce Fig.~\ref{fig:xsec_vs_Ev}, but now we demonstrate the individual contributions of positive or negative parities for the given $J$-transitions. The results are shown in Fig.~\ref{fig:xsec_vs_Ev_parities} for both $^{203,205}$Tl isotopes. For details see Sec.~\ref{sec:results}.

\begin{figure}
    \centering
    \includegraphics[width=\textwidth]{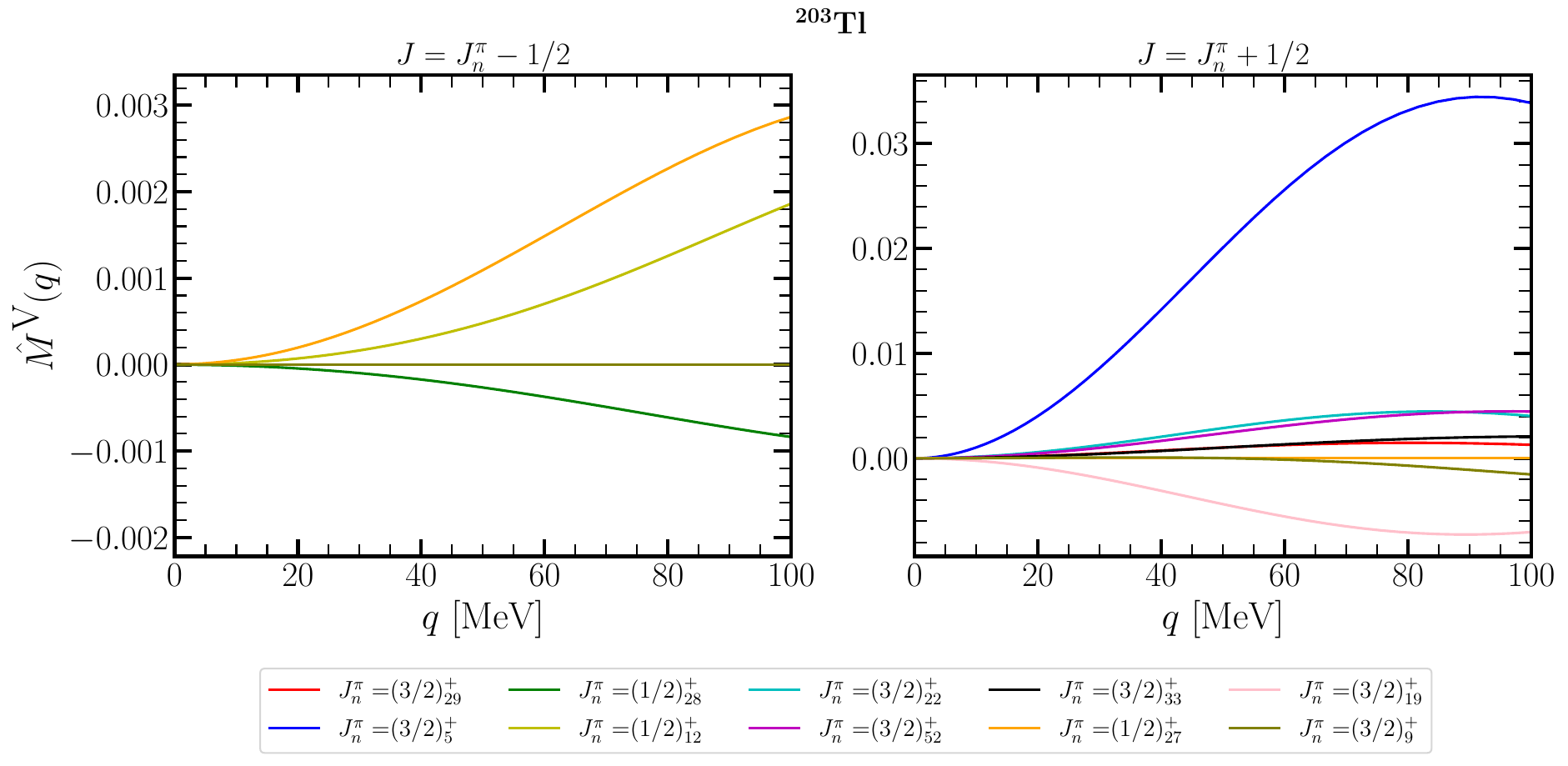}
    \includegraphics[width=\textwidth]{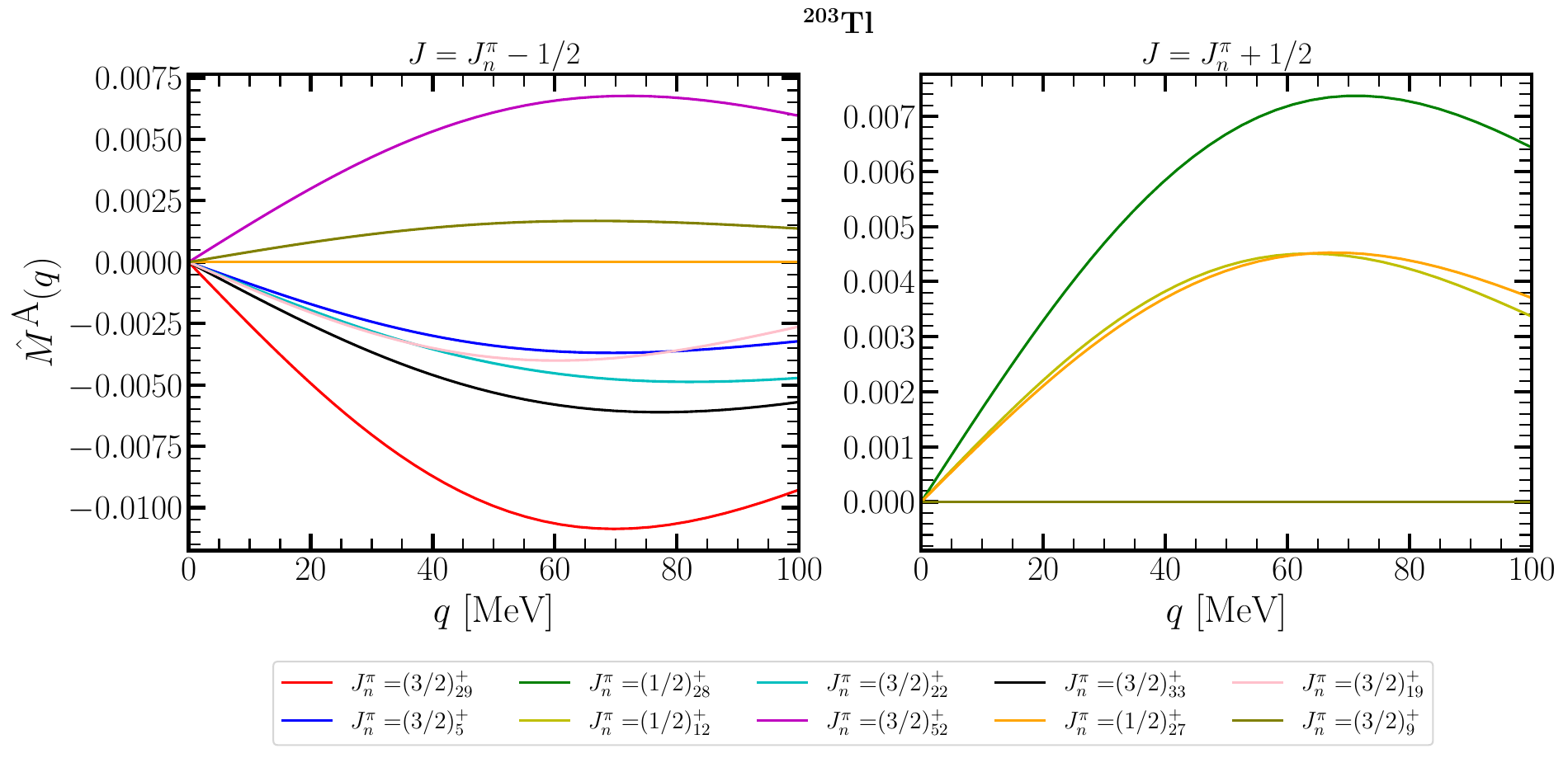}
   
    \caption{The Coulomb matrix elements as functions of the three-momentum transfer $q$ for $^{203}$Tl. Upper (lower) panels show the $\hat{M}^{\textrm{V}}$ ($\hat{M}^{\textrm{A}}$) component of $\hat{\mathcal{M}}$. The results are presented for transitions to the first ten most contributing final nuclear states of the inelastic $^8$B neutrino-$^{203}$Tl scattering cross section from the ground state to the final state $J^
    \pi_n$, i.e. from $1/2^+_1 \to J^\pi_n$. Left (right) panels correspond to the transitions with $J=J^{\pi}_n-1/2$ ($J=J^{\pi}_n+1/2$). For details see the text.}
    \label{fig:ME_coulomb_Tl203}
\end{figure}

\begin{figure}
    \centering
    \includegraphics[width=\textwidth]{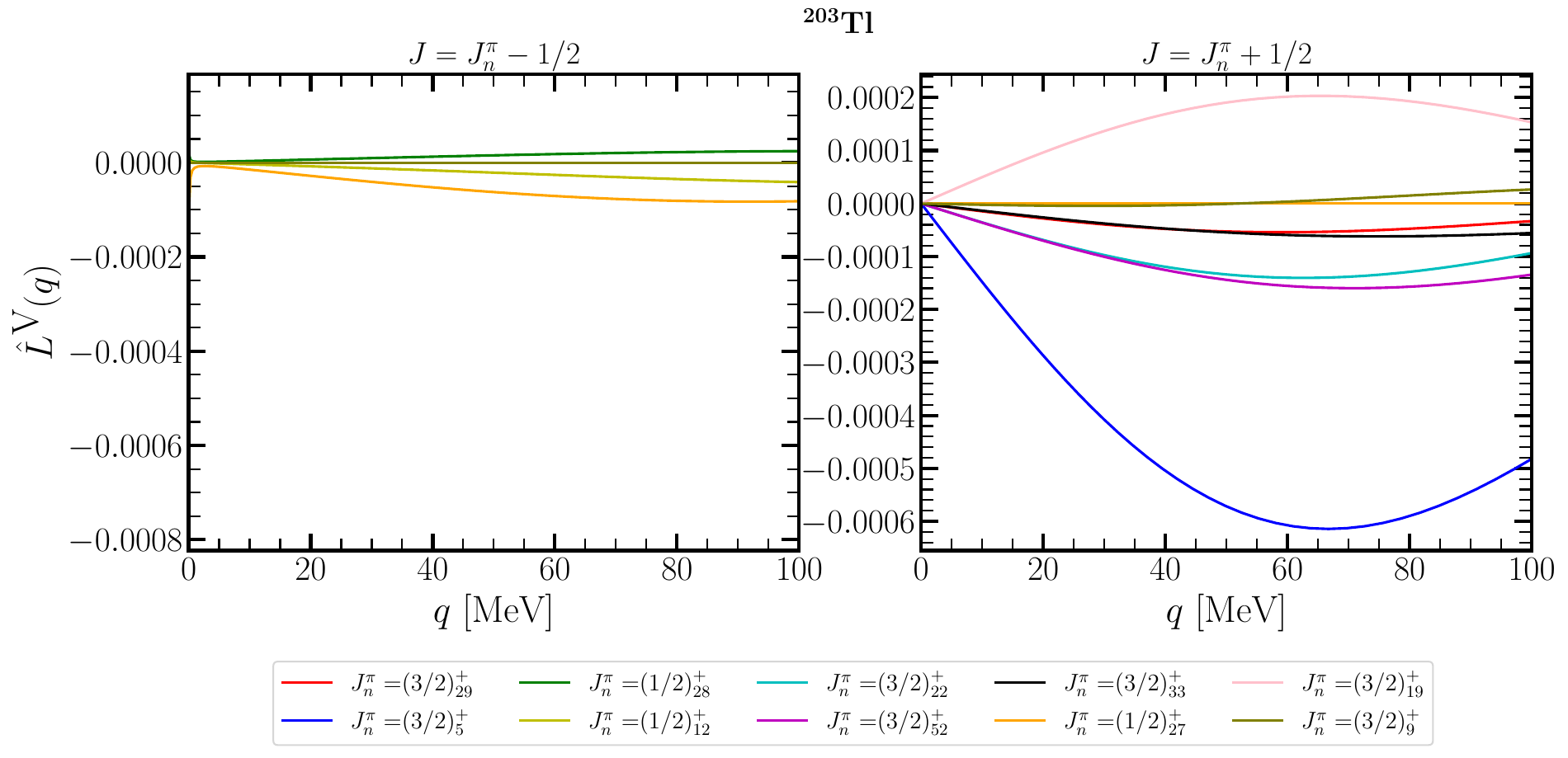}
    \includegraphics[width=\textwidth]{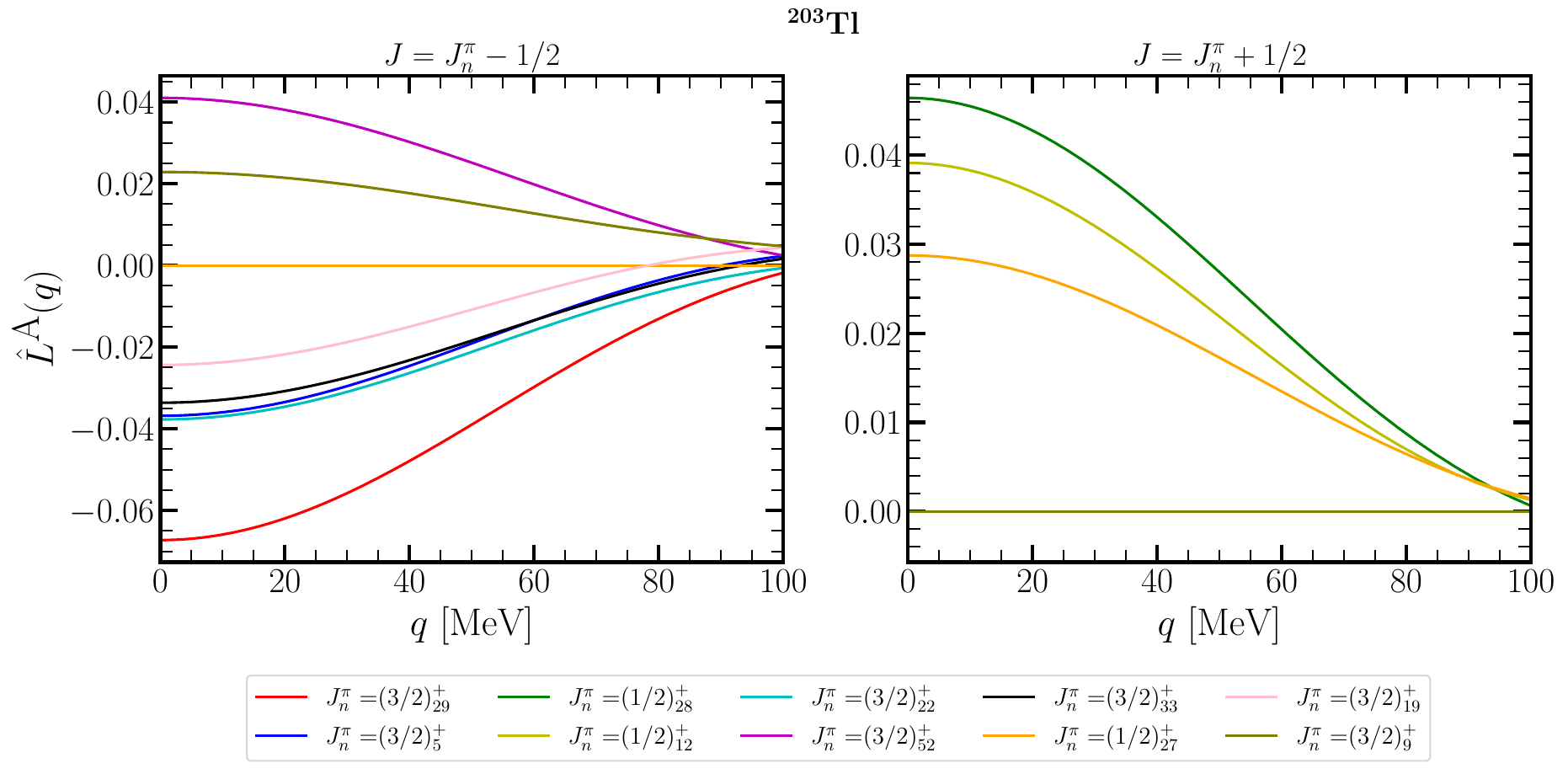}
   \caption{Same as Fig.~\ref{fig:ME_coulomb_Tl203} but for the longitudinal matrix elements $\hat{L}^{\textrm{V}}$ and $\hat{L}^{\textrm{A}}$.}
    \label{fig:ME_longitudinal_Tl203}
\end{figure}

\begin{figure}
    \centering
    \includegraphics[width= \textwidth]{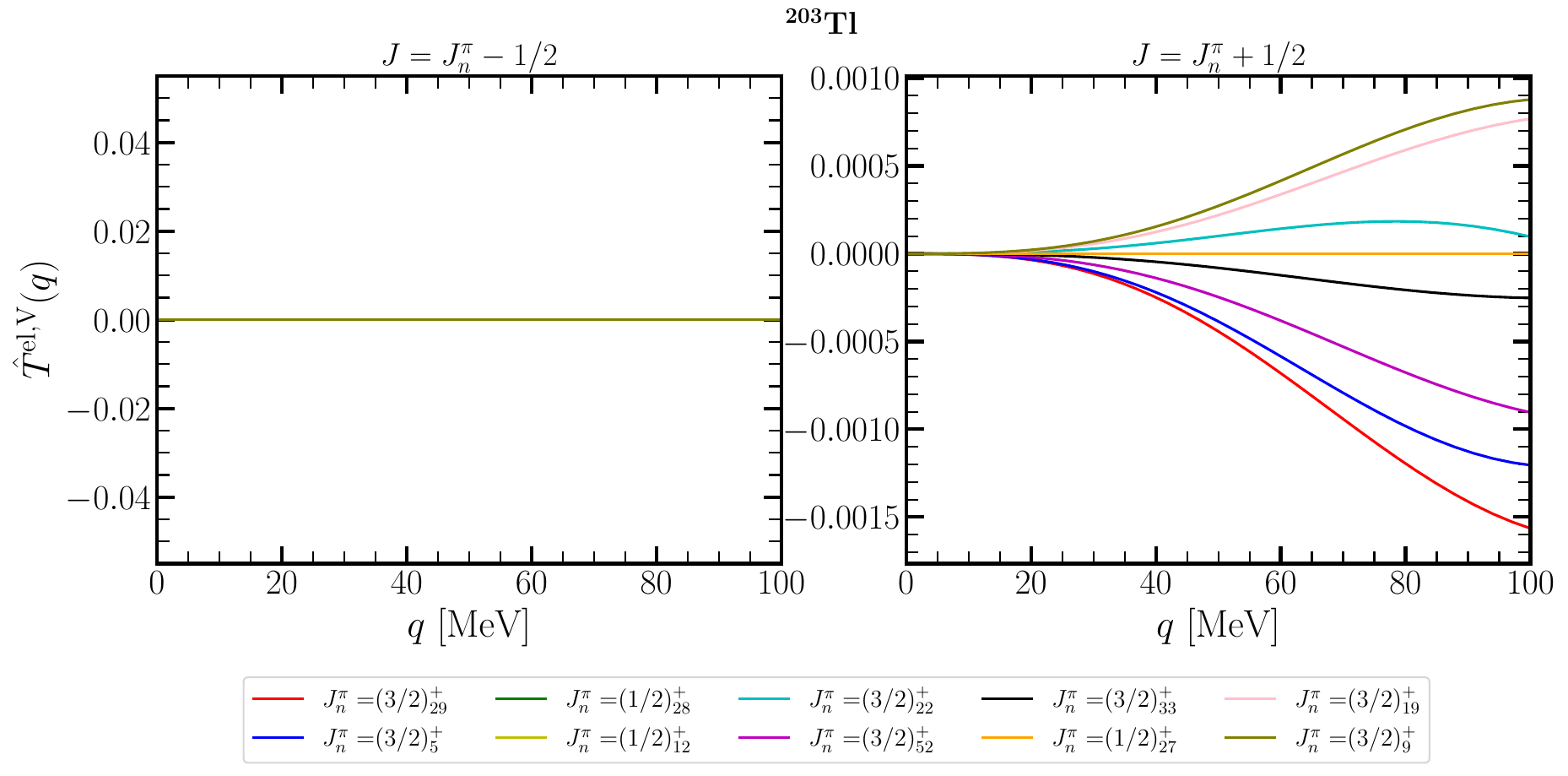}
    \includegraphics[width= \textwidth]{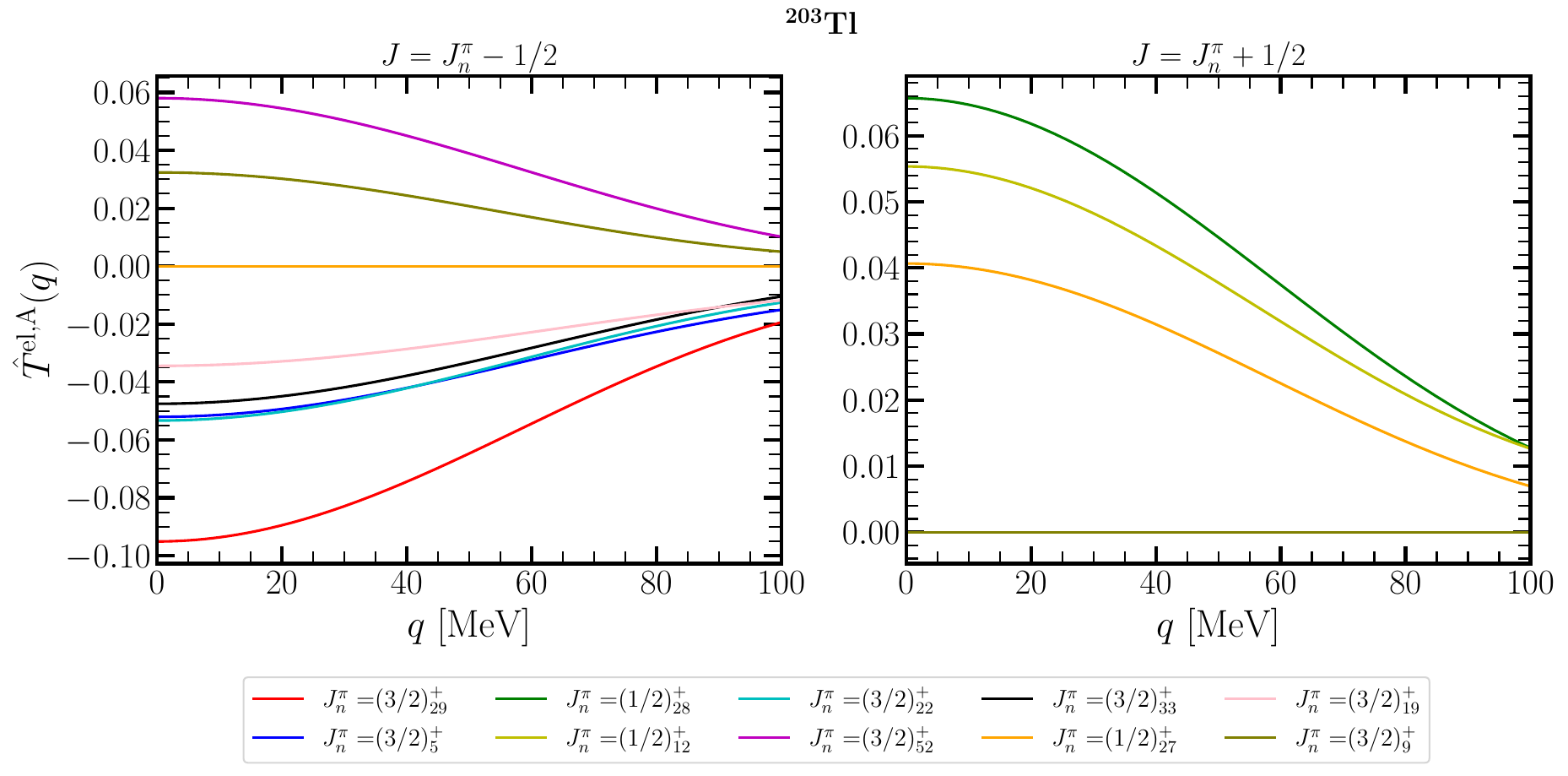}

    \caption{Same as Fig.~\ref{fig:ME_coulomb_Tl203} but for the transverse electric matrix elements $\hat{T}^\mathrm{el, V}$ and $\hat{T}^\mathrm{el, A}$.}
    \label{fig:ME_Tel_Tl203}
\end{figure}

\begin{figure}
    \centering
    \includegraphics[width= \textwidth]{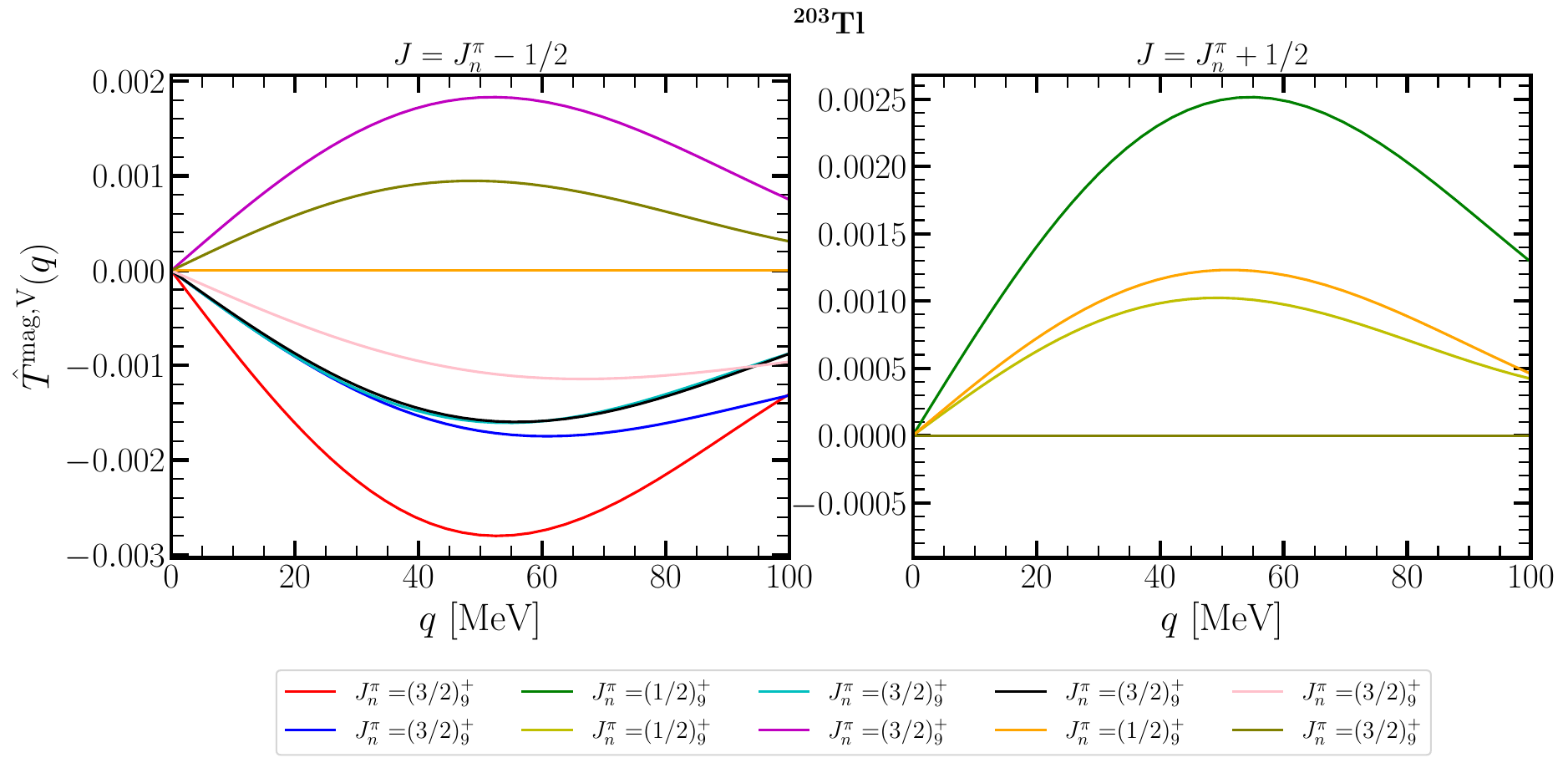}
    \includegraphics[width= \textwidth]{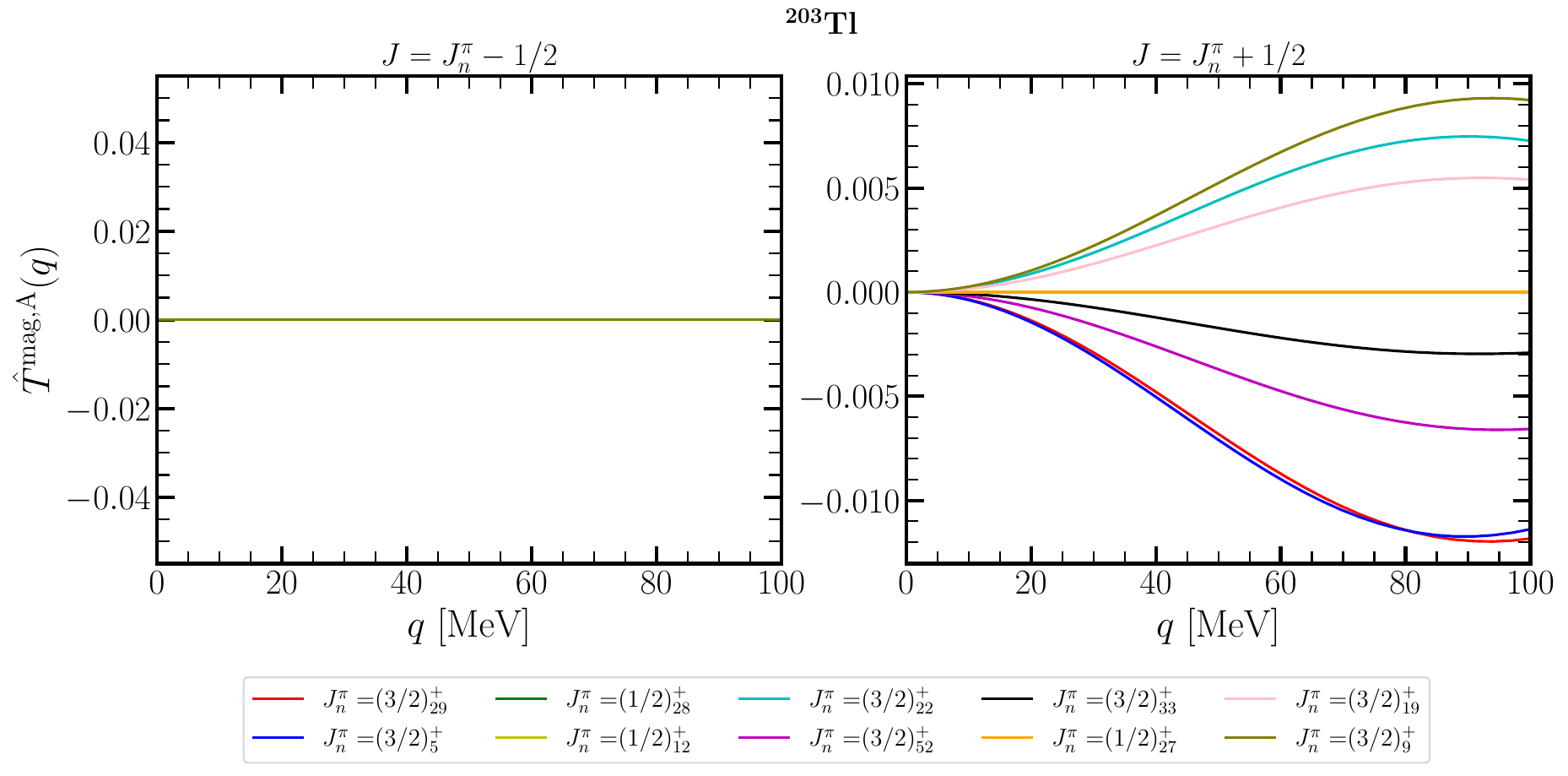}
    \caption{Same as Fig.~\ref{fig:ME_coulomb_Tl203} but for the transverse magnetic matrix elements $\hat{T}^\mathrm{mag, V}$ and $\hat{T}^\mathrm{mag, A}$.}
    \label{fig:ME_Tmag_Tl203}
\end{figure}

We also provide the matrix elements of $\hat{\mathcal{M}}_J$, $\hat{\mathcal{L}}_J$, $\hat{\mathcal{T}}^\text{el}_J$, and $\hat{\mathcal{T}}^\text{mag}_J$, for both vector and axial-vector components, as functions of the three-momentum transfer. The respective results are given in Figs.~\ref{fig:ME_coulomb_Tl203}--\ref{fig:ME_Tmag_Tl203} for the case of $^{203}$Tl and correspond to the $J$-transitions to the ten most contributing final nuclear states in $^8$B neutrino scattering. Although not shown here, the matrix elements of $^{205}$Tl are rather similar. The matrix elements are given for transitions between the initial nuclear state (ground state) of multipolarity $J^\pi_\mathrm{g.s.} = 1/2^+_1$ to final nuclear states denoted as $J^\pi_n$, where $\pi=\pm$ denotes the parity and $n$ enumerates the final states. The left (right) panels correspond to the $J = J^\pi_n - 1/2$ ($J = J^\pi_n + 1/2$) transition in each case, by noting that since the nuclear ground  state is $1/2^+$ there are only two possible transitions (e.g., the red plot on the top left panel of Fig.~\ref{fig:ME_coulomb_Tl203} corresponds to the matrix element of the $M^{\textrm{V}}_J(q)$ operator of a transition with $J^{\pi} = 1^+$ from the g.s. to the excited state $J^{\pi}_n = 3/2^+_{29}$, and the same graph on the top right panel corresponds to the same matrix element for the $J^{\pi} = 2^+$ transition). For $^{203}$Tl, the first ten most contributing states in descending order are $1/2^+_1 \to J^\pi_n=$
 \{$(3/2)^+_{29}$,
$(3/2)^+_{5}$,
$(1/2)^+_{28}$,
$(1/2)^+_{12}$,
$(3/2)^+_{22}$,
$(3/2)^+_{52}$,
$(3/2)^+_{33}$,
$(1/2)^+_{27}$,
$(3/2)^+_{19}$,
$(3/2)^+_{9}$\}.
Similarly for $^{205}$Tl, the first ten most contributing states in descending order are: $1/2^+_1 \to J^\pi_n=$
 \{$(3/2)^+_{7}$,
$(3/2)^+_{3}$,
$(1/2)^+_{6}$,
$(3/2)^+_{19}$,
$(1/2)^+_{11}$,
$(1/2)^+_{9}$,
$(3/2)^+_{2}$,
$(1/2)^+_{3}$,
$(1/2)^+_{10}$,
$(1/2)^+_{7}$\}. The interested reader is referred to Fig.~\ref{fig:contributions-final-states-8B}.

There are a number of points worth discussing regarding the plots of the matrix elements. Firstly, certain transitions vanish entirely. This is due to the selection rules of the multipole operators. The operators $\hat{M}^{\textrm{V}}_{J}$, $\hat{L}^{\textrm{V}}_{J}$, $\hat{T}^{\textrm{el, V}}_{J}$, and $\hat{T}^{\textrm{mag, A}}_{J}$ have parity $(-1)^J$ whereas the operators $\hat{M}^{\textrm{A}}_{J}$, $\hat{L}^{\textrm{A}}_{J}$, $\hat{T}^{\textrm{el, A}}_{J}$, and $\hat{T}^{\textrm{mag, V}}_{J}$ have parity $(-1)^{J+1}$. Because of this, either the transition with $J = J_{n}^{\pi} - 1/2$ or the one with $J = J_{n}^{\pi} + 1/2$ has to necessarily vanish since they cannot both fulfill the selection rule. Whether the higher or the lower $J$ transition vanishes is entirely determined by the final state $J^{\pi}_n$ angular momentum and parity.

Second, for low 3-momentum transfers only the $\hat{L}_1^{\textrm{A}}$ and $\hat{T}_1^{\textrm{el, A}}$ operators produce non-vanishing contributions. This is just the allowed limit where these two operators are proportional to the Gamow-Teller operator and the identity $\hat{\mathcal{T}}_1^{\textrm{el}} = \sqrt{2}\hat{\mathcal{L}}_1$ holds. As the states considered were selected based on their contribution to the relatively low energy $^{8}$B neutrino scattering, their order from the most contributing to the least contributing can be readily seen by the matrix elements of these two operators at low $q$ values. As the transferred 3-momentum increases the matrix elements of these two operators decrease, while those of the other six which obey selection rules increase. The behavior of the matrix elements as a function of $q$ is similar for both nuclei of interest, and we have thus included figures for only $^{203}$Tl.

\end{document}